\documentclass[journal]{IEEEtran}

\usepackage{epsfig}
\usepackage{graphicx}
\usepackage{amsmath}
\usepackage{amssymb}
\usepackage{color}
\usepackage{mathtools}
\usepackage[capitalise]{cleveref}

\usepackage{subcaption}

\usepackage{cite}

\ifCLASSINFOpdf
\else
\fi
\usepackage{url}

\begin{document}

\title{Controlling Rayleigh-Backscattering-Induced Distortion in Radio over Fiber Systems for Radioastronomic Applications}


\author{Jacopo~Nanni, 
		Andrea~Giovannini, 
		Muhammad~Usman~Hadi, 
		Enrico~Lenzi, 
		Simone~Rusticelli,\\  
		Randall Wayth  ˜\IEEEmembership{Member, ˜IEEE,}
		Federico~Perini, 
		Jader~Monari, 
		Giovanni~Tartarini  ˜\IEEEmembership{Member, ˜IEEE} 
\thanks{J.~Nanni, A.~Giovannini, M.~U.~Hadi and G.~Tartarini are with the Dipartimento di Ingegneria dell'Energia Elettrica e dell'Informazione ``Guglielmo Marconi'', Universit\`a di Bologna, 40136 Bologna (BO), Italy (e-mail: jacopo.nanni3@unibo.it; andrea.giovannini10@studio.unibo.it; muhammadusman.hadi2@unibo.it; giovanni.tartarini@unibo.it).}
\thanks{F.~Perini, J.~Monari and S.~Rusticelli are with Institute of Radio Astronomy, National Institute for Astrophysics, Via Fiorentina 3513, 40059 Medicina (BO), Italy (e-mail: f.perini@ira.inaf.it;j.monari@ira.inaf.it;rusticel@ira.inaf.it).}
\thanks{E.~Lenzi is with Protech S.a.S, Via dei Pini 21, 31033 Castelfranco Veneto (TV), Italy (e-mail:e.lenzi@protechgroup.it).}
\thanks{R.~Wayth is with Curtin University, Bentley
Western Australia, 6102 (e-mail:R.Wayth@curtin.edu.au).}}

\maketitle

\begin{abstract}
Radio over Fiber (RoF) Systems exploiting a direct modulation of the laser source are presently utilized within important Radioastronomic scenarios. Due to the particular operating conditions of some of these realizations, the phenomena which typically generate nonlinearities in RoF links for telecommunications applications can be here regarded as substantially harmless. However, these same operating conditions can make the RoF systems vulnerable to different kinds of nonlinear effects, related to the influence of the Rayleigh Backscattered signal on the transmitted one. A rigorous description of the phenomenon is performed, and an effective countermeasure to the problem is proposed and demonstrated, both theoretically and experimentally. 
\end{abstract}


\section{Introduction}
\IEEEPARstart{T}{he} transmission of radio frequency (RF) signals over the optical fiber through the so-called Radio-over Fiber (RoF) technology is nowadays part of many infrastructures related to applications which include telecommunications (4G/5G signals distribution, CATV, etc..), monitoring and radioastronomic signals reception \cite{Raweshidy,Dat_Kanno,Urban,Weiss}. Within these scenarios the advantages of optical fibers with respect to coaxial cables are exploited to reach higher quality of transmission (such as  lower attenuation and higher bandwidth), lower cost, lower dimensions and electromagnetic interference immunity. 

For those of the aforementioned applications which are cost sensitive, the RoF systems design can be characterized by the use of the simple analog transmission, realizing a direct intensity modulation of the laser source and a direct detection at the receiver's end (D-IMDD scheme). This is the case of significant telecommunications scenarios \cite{Analog_Rof_mm_Wave_2020,MWP18,JQE19} as well as important radioastronomic ones like downlinks of large dishes belonging to Very Large Baseline Interferometry (VLBI) Networks \cite{YEBES}, or the downlinks of the planned 130,000 receiving antennas of low RF frequency detection system within the Square Kilometre Array (SKA) project (SKA-LOW \cite{SKA_LOW}). 

{For all these possible RoF applications, one of the major issues to be faced is the insurgence of nonlinearities in the input-output characteristic of the system, which determines the undesired creation of spurious frequency components at the receiver side placed at the fiber end.}

{In case of RoF systems operating within the mobile network, the spurious frequency terms result from the interaction among the spectral components of the signals transmitted in the fiber, bringing detriment, for a given signal,  both within its own bandwidth and in the bandwidth occupied by other signals sharing the same transmission channel \cite{cox}. Within this context, if the fiber utilized is of multimodal type, one major cause of nonlinearities is given by the presence of the phenomenon of Modal Noise \cite{Alcaro}. This phenomenon disappears if single mode fibers are utilized, in which case nonlinearities are typically caused by the combined action of laser chirp and fiber chromatic dispersion \cite{Chirp_Dispersion_2010}, or by the non perfect linearity of the slope-efficiency curve exhibited by the laser source \cite{Petermann}. In these last cases  countermeasures must be taken, like the adoption of appropriate digital predistortion schemes \cite{OQE_Usman,ASTESJ}.}

{When RoF systems are utilized within Radioastronomic scenarios, the signals coming from sky sources and traveling in the fiber typically exhibit very low levels of both power and coherence, and do not consequently give rise to the generation of spurious frequency components. Together with the sky signals it has however in this case to be considered the presence of the so called Radio Frequency Interference signals (RFIs).} {Excluding those due to lightning from nearby thunderstorms, RFIs come from known coherent sources (e.g. high power FM transmitters located within a few hundreds of km, artificial satellites, radio and television signals reaching the site due to tropospheric ducting),} { and can be considered as sinusoidal tones whose frequency and maximum amplitude are regularly registered and monitored. Although their power levels at the laser input of the RoF link are much smaller than the typical powers of the input RF signals in RoF telecommunication systems, in the vicinity of their respective frequencies RFIs exhibit a power that is much higher than that of the sky signal. It is therefore very important that the RoF links realized do not exhibit nonlinear characteristics,  since this would imply the generation of higher harmonics as well as inter-modulation products of these RFIs, which can have a detrimental effect on the desired quality of the received signals.}

\indent The possible causes of nonlinearity of the RoF link listed above can actually be eliminated or reduced to negligible levels, in case of Radioastronomic applications. Indeed, choosing to operate with G.652 single mode fibers, rules out the insurgence of Modal Noise in case the optical wavelength utilized is above 1260nm \cite{Fotonica}. Moreover, operating in the second optical window, i.e. around 1310nm, which guarantees zero chromatic dispersion maintaining at the same time acceptable attenuation (due to the distances involved which are typically lower than 10km), reduces to negligible levels the combined effect of laser chirp and fiber chromatic dispersion. Finally, given the fact that even after preamplification the RFIs tones powers reaching the RoF links  reach levels around $-20\,dB_m$ or less,  the level of the spurious frequencies due to the imperfection of the laser slope-efficiency characteristic typically falls below the noise floor.

Other causes of nonlinearity could be given by the interaction of the transmitted signal with portions of itself which are reflected, e.g. in correspondence to imperfect connectors \cite{Lidgard}, or scattered, due to the Brillouin effect \cite{Yoshinaga,Peral}. However, D-IMDD RoF systems for Radioastronomic applications, like all modern optical communications systems, are furnished by Angle Polished Connectors (APC), which provide return losses of the order of $60$ dB, and avoid the impairments due to connectors reflections. Moreover, their already cited typical length values, together with their corresponding launched optical powers which are typically not higher than $4-6\,dB_m$, exclude appreciable impairments due to the presence of Brillouin scattering \cite{Feng}.

One cause of nonlinearities which in this scenario cannot be {\it a priori} neglected consists instead in the interaction between the optical signal and its portion which, after having been backscattered by Rayleigh effect, is reflected at the transmitter's section of the link and propagates to the receiver's end. The effects of Rayleigh Backscattering (RB) in optical fiber systems have been extensively studied, with a substantial focus on the undesired reduction of the Signal to Noise Ratio at the receiver's end {in optical fiber systems for telecommunications applications.} \cite{Wu,Wan,Gysel_JLT}. 

However, the insurgence of nonlinearities related to RB in D-IMDD RoF links has been put into evidence in relatively recent times \cite{ICTON_SKA}. The reason consists in the fact that its detriments can become important only in presence {of particular combinations of the operating parameters of the optical system considered, which include e.g. amplitude and frequency of the modulating signal, combined with the constraint of maintaining at low levels the total cost of the whole RoF link.  While only in particular cases these operating conditions and constraint can be appreciated in typical optical systems for telecommunications, in case of applications like the Radioastronomic ones, this framework  constitutes a possible normal operating condition, and the detriments due to RB-induced nonlinearities have then to be adequately characterized and counteracted.}

In the present paper the onset of undesired nonlinear characteristics in {D}-IMDD RoF systems will be described in detail through the derivation of a rigorous mathematical model, which will be validated through accurate comparisons with experimental results. The analysis will allow to predict the insurgence of the problem in RoF systems designed for Radioastronomic applications. Extending the work presented in \cite{MWP2019}, an effective solution to reduce to acceptable levels the nonlinearities related to RB  will subsequently be proposed, and analyzed, {arriving to identify some important} design parameters. 

The paper is organized as follows. {In Section II the mathematical model which describes the insurgence of nonlinearities related to RB in D-IMDD RoF systems will be developed. {In Section III it will be evidenced how their impact is of particular importance in the context of some radioastronomic applications}. In section IV the same developed model will be utilized to propose an effective countermeasure to the problem illustrated. In Section V the model will be successfully tested through a comparison between measured and simulated behaviors of quantities related to system nonlinearities. The experimental test will regard also the solution adopted, whose effectiveness will be confirmed, leading to the optimization of some of its operating parameters.} Finally conclusions will be drawn.





\section{Theoretical description of the undesired effect}
\label{sec:theoretical}
\subsection{{Electrical field emitted by the Laser and coupled into the Optical Fiber}}
The optical link considered is composed of a DFB laser source, connected to a span of Standard Single Mode Fiber (SSMF), directly connected to a PIN photo-detector. The laser, having respectively threshold and bias currents $I_{th}$ and $I_{bias}$, is modulated by a current composed of $N_s$ radio frequency sinusoidal signals expressed by:
\begin{equation}
 I_{RF}(t)=\sum_{i=1}^{N_s} I_{RF,i,0}\cos(\omega_{RF,i}t)   
\end{equation}
in which $I_{RF,i,0}$ is the amplitude of the $i$--th signal and where the angular frequencies $\omega_{RF,i}|_{i=1,\dots,N_s}$, {all exhibit values which are well below the relaxation resonance angular frequency of the laser source utilized}.

The {difference between the total current injected into the laser $I_{inj}(t)=I_{bias}+I_{RF}(t)$ and the threshold current $I_{th}$} can then be conveniently put in the form:

{\begin{eqnarray}\label{eq:I_RF}
&&I_{inj}(t)-I_{th} = (I_{bias}-I_{th}) \left[1+ i_{RF}(t) \right] =\nonumber \\ 
&&= (I_{bias}-I_{th}) \left[1+ \sum_{i=1}^{N_s} OMI_i\cos(\omega_{RF,i}t) \right]
\end{eqnarray}}
{where the {\it Optical Modulation Index} (named also {\it intensity modulation index}) of the $i$-th signal $OMI_i = I_{RF,i,0}/(I_{bias}-I_{th})$ has been introduced}

{According to the  physical process described by the Laser Rate Equations, the injection current $I_{inj}(t)$, influences the variation in time of the number $N_c(t)$ of charge carriers in the laser active region. This, in turn, determines the time behavior of the number of photons in the active region, given by:
\begin{eqnarray}
&&N_p(t)=\eta_p (I_{bias}-I_{th})\left[1+i_{RF}(t) +a_2 (i_{RF}(t))^2+\right. \nonumber \\
&& \left. + a_3 (i_{RF}(t))^3+... \right]=\eta_p (I_{bias}-I_{th})[1+s(t)]
\label{eq:N_p}
\end{eqnarray} 
where $\eta_p$ is the linear component of the Photon-Current laser characteristic curve (considered to be the same for $I_{bias}-I_{th}$ and for $I_{RF}(t)$), while $a_2$ and $a_3$ account for second and third order nonlinearities of the same curve, respectively.} 

{Following the approach proposed, e.g. in \cite{Petermann}, \cite{Agrawal}, a complex representation of the electrical field $\vec{E}_{TX}(t)$ generated inside the laser cavity and coupled to the single mode optical fiber can be introduced. This field oscillates in the vicinity of the angular frequency $\omega_{th}$ correspondent to the lasing threshold, and exhibits an amplitude $A(t)$ and and phase $\zeta(t)$ which can be expressed as:}

{
\begin{eqnarray}
&&\vec{E}_{TX}(t) = A(t) e^{j \zeta(t)}e^{j \omega_{th} t} \vec{e}_{01}  =  
A(t) e^{j\left[\theta_0(t)+\phi(t)\right]} \times \nonumber \\ 
&&\times e^{j \omega_{th} t} \vec{e}_{01} = A_0\sqrt{N_p(t)}e^{j \left[K \int N_p(t) +\phi(t)  \right]} e^{j\omega_{th}t} \vec{e}_{01}
\label{eq:E_a}
\end{eqnarray}}

{from which it results that both $A(t)$ and $\zeta(t)$  are functions of $N_p(t)$. The expression of $A(t)$ indicates the presence of the laser intensity modulation. In particular, the constant quantity $A_0$ is such that $A_0^2$ represents the proportionality constant between $N_p(t)$ and the optical power  emitted by the laser and coupled into the optical fiber. This implies that the dimensions of $E_{TX}(t)$ are actually $\frac{V/m}{\sqrt{\Omega}}$ (while in \cite{Agrawal} they are $V/m$ and in \cite{Petermann} the field is adimensional), and this will be maintained in the following for formal simplicity of the mathematical derivation.}

{Moreover,  the phase  $\zeta(t)$ of the Electrical Field, besides the laser phase noise contribution $\phi(t)$, exhibits a term $\theta_0(t)= K \int N_p(t)$ which denotes the presence of a phase (or frequency) modulation of the optical field called \emph{frequency chirp} \cite{Agrawal}, generated by the direct modulation of the laser. Note that, due to the values assumed by  $\omega_{RF,i}|_{i=1,\dots,N_s}$, in the expression of $\zeta(t)$ the transient chirp term could be neglected  \cite{Petermann}. Note also that the value of the coefficient $K$, which accounts for both adiabatic and thermal chirp effects, can vary with the frequency of the term it multiplies.} 
{Finally, in Eq. \eqref{eq:E_a} $\vec{e}_{01}$ is the normalized mode function referred to the fundamental $LP_{01}$ mode.}

{As a consequence of the considerations developed above, the expression of $\vec{E}_{TX}(t)$ can be rewritten as:}

\begin{equation}
\vec{E}_{TX}(t) = E_0\sqrt{1+s(t)}\cdot e^{j[\theta(t)+\phi(t)]}e^{j\omega_0t}\vec{e}_{01}
\label{eq:E_TX}
\end{equation}

where $E_0=\sqrt{P_0}$ is the electrical field amplitude, with $P_0=\eta (I_{bias}-I_{th})$ representing the optical power coupled into the optical fiber if only the bias current were injected, and where $\eta$ is the Power-Current laser slope efficiency. {Moreover,  it is $\theta(t)=\theta_0(t)-K\eta_P(I_{bias}-I_{th})t$, while $\omega_0 =$ $\omega_{th}+K\eta_P(I_{bias}-I_{th})$ $= 2\pi f_0$ is the optical frequency of emission}.

{Putting, for simplicity of notation, $K\eta_P=K_f(\omega_{RF})$, in the considered case of} modulation performed by simple sinusoidal signals of angular frequency $\omega_{RF,i}$, a corresponding phase modulation index can defined as $M(\omega_{RF,i})=2\pi K_f(\omega_{RF,i})I_{RF,i,0}/\omega_{RF,i}$ \cite{Nanni_AO}. The phase modulation  $\theta(t)$ due to the frequency chirp  can then be expressed as:

\begin{equation}\label{eq:theta}
\theta(t) = \sum_{i=1}^{N_s}M(\omega_{RF,i})\sin(\omega_{RF,i}t)
\end{equation}

The expression of the field after a length $z$ can be expressed as follows:

{
\begin{align}\label{eq:campo_no_scattering}
&\vec{E}_{TX}(t,z) = E_0\sqrt{1+s(t-\hat\tau z)} \times \nonumber\\
&e^{j[\theta(t-\hat\tau z)+\phi(t-\hat\tau z)]}e^{j(\omega_0t - \beta z)-\frac{\alpha}{2} z}\vec{e}_{01}
\end{align} }

where $\hat\tau$ is the group delay-per-meter of the fundamental mode, {$\beta$ is its propagation constant}, while $\alpha$ is the attenuation coefficient of the fiber material at the optical frequency considered in $neper/m$. 
\subsection{Description of the RB signal}
Because of the imperfections on the refractive index along the optical fiber, part of the optical signal is scattered, in both forward and backward directions. As mentioned in the Introduction, this phenomenon is called \emph{Rayleigh Backscattering} (RB) and because of its nature results to be elastic and therefore linear. The total amount of RB can be seen as the sum of the contributions of all the scatterers present along the optical path. 

\begin{figure}[hbtp]
\centering
\includegraphics[scale=0.23]{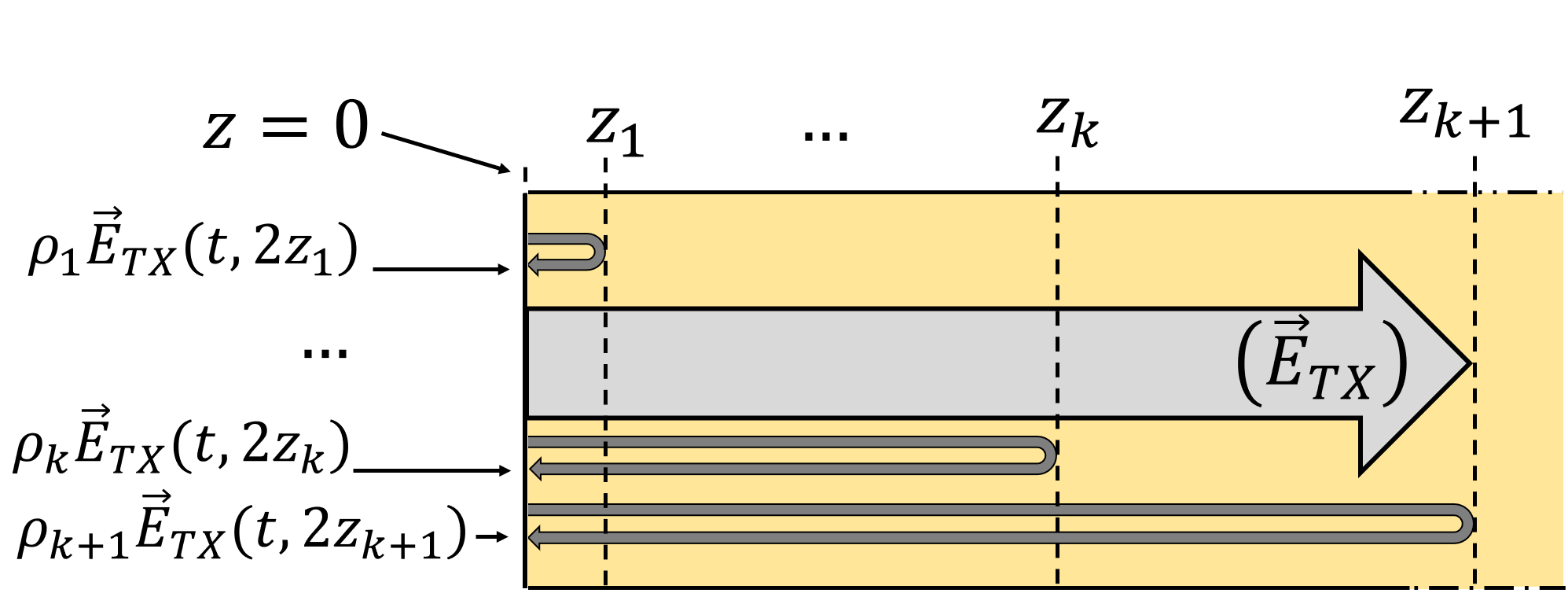}
\caption{Scheme of the Rayleigh Backscattering {(RB)} model considered. See text for details.}
\label{fig:Rayleigh_principle}
\end{figure}

Figure \ref{fig:Rayleigh_principle} shows schematically the concept of the RB model considered. 

In particular, at the fiber input section ($z=0$) 
the total backscattered field $\vec{E}_{RB}(t,z=0)$ is given by the sum: 

\begin{equation}\label{eq_E_RB}
\vec{E}_{RB}(t,z=0) = \sum_{k=1}^{k_{max}}\rho_k\vec{E}_{TX}(t,2 z_k)
\end{equation} 

 where $z_k$ is the coordinate of the generic scattering section and $k_{max}$ is the total number of scattering sections (i.e. $z_{k_{max}}=L$, where $L$ is the total length of the fiber span). Moreover, $\rho_k=\rho(z_k)$ is the reflection coefficient at section $z_k$, while the term $2z_k$ in the argument of $E_{TX}$ takes into account the round-trip of the scattered field. 

Following the model proposed in  {\cite{Gysel_JLT}}, the quantity {$\rho_k$ is assumed to be a time independent zero-mean complex Gaussian variable. The variance of both its real and imaginary parts is $\sigma^2_{\Re{\bf e}\{\rho_k\}/\Im{\bf m}\{\rho_k\}}=\frac{1}{2}\alpha_s\mathcal{S}z_k$} where $\alpha_s$ is the Rayleigh attenuation coefficient, which for the considered wavelengths can be assumed to coincide with $\alpha$, while $\mathcal{S}$ is the so-called \emph{backscattering factor} or \emph{recapture factor}{ \cite{Brinkmeyer}}, which depends on the characteristics of the fiber considered, exhibiting typical values of the order of $10^{-3}$ for the standard G652 fiber.\\ {The quantities $\rho_k$ are also assumed to be {\it delta correlated} as follows:} 
\begin{equation}\label{eq:rho_k}
    {\mathbb{E}[\rho_k\rho^*_h]= \left\{\begin{array}{ll}
    \sigma^2_{\rho_k}=2\sigma^2_{\Re{\bf e}\{\rho_k\}/\Im\{\rho_k\}} & if \quad k = h\\
    \\
    0 &if \quad k\neq h
    \end{array}\right.}
\end{equation}
{\noindent where $\mathbb{E}[\cdot]$ represents the expected value operator}. \\
\indent The corresponding average optical power back scattered at the input section $P_{RB}(z=0)$ can be shown \cite{Gysel_JLT} to be given by:

\begin{equation}
P_{RB}(z=0) = P_0\frac{\alpha_s\mathcal{S}(1-e^{-2\alpha L})}{2\alpha}\simeq P_0\frac{\mathcal{S}(1-e^{-2\alpha L})}{2}
\end{equation}

\subsection{Modeling of the nonlinearities after the RB laser feedback}

\begin{figure}[hbtp]
\centering
\includegraphics[scale=0.28]{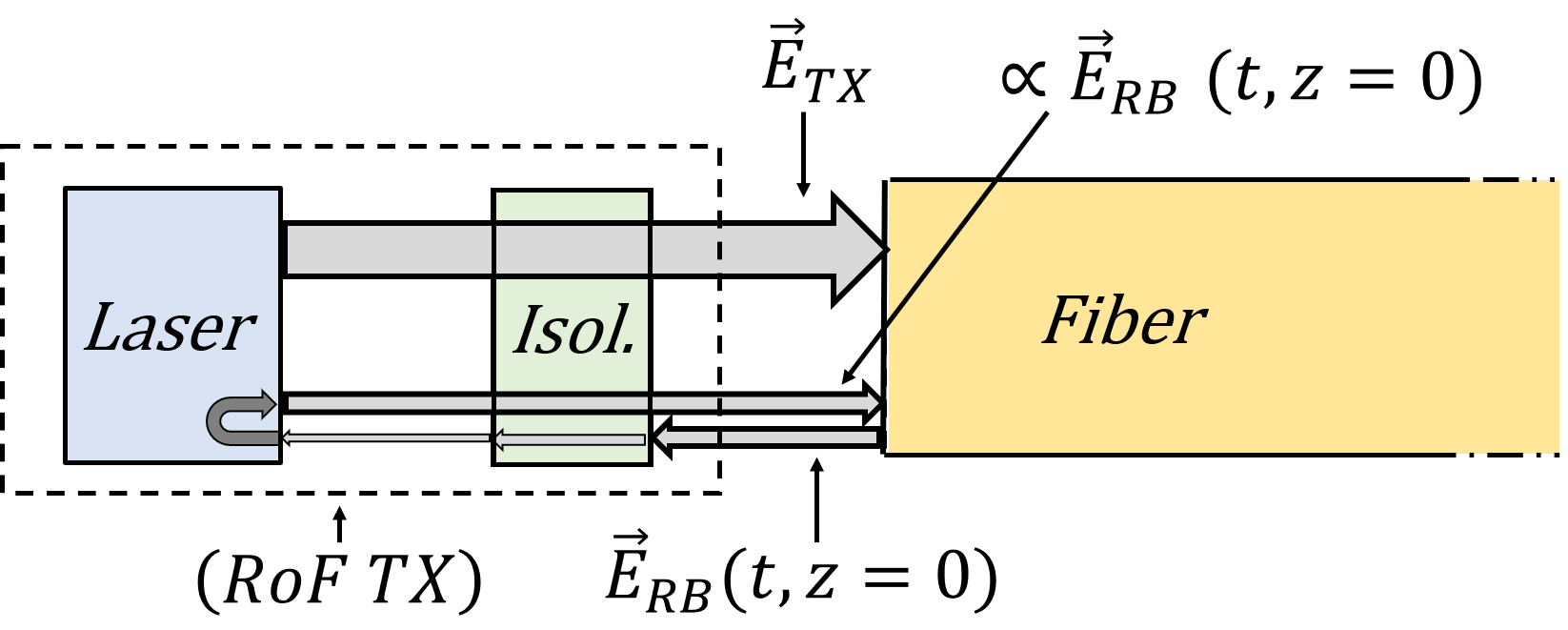}
\caption{Generation at the input section of a fiber of a replica of ${E}_{RB}(t,z=0)$ propagating in the positive $z$ direction.}
\label{fig:Laser_feedback}
\end{figure}

Despite the presence of an optical isolator, which is part of the RoF Transmitter (RoF TX) and attenuates the backreflection by typically 30-40 dB, a small portion of $P_{RB}(z=0)$ (or, equivalently, of ${E}_{RB}(t,z=0)$) results to be fed back into the laser source. This determines \cite{Goldberg,Chraplyvy,Gysel_JLT}, among other effects, { the re-emission by the laser itself  of a field proportional to $\vec{E}_{RB}$ through a coefficient $\Gamma = \sqrt{\frac{G_L}{Att_{iso}}}$, being $G_L$ and $Att_{iso}$ the laser amplification factor on the reflected signals and the isolator power attenuation, respectively.}
Figure \ref{fig:Laser_feedback} illustrates the process that has just been described.

This field finally reaches the photodetector placed at the coordinate $z=L$ together with the transmitted signal $\vec{E}_{TX}$. The current generated by the photo-detector can then be computed as:
\begin{eqnarray}
&i_{out}(t)&=\mathcal{R}\left|\vec{E}_{TX}(t,L) + \vec{E}_{RB}(t,L)\right|^2\nonumber\\
&&=\mathcal{R}\left[\left|\vec{E}_{TX}(t,L)\right|^2+\left|\vec{E}_{RB}(t,L)\right|^2+\right.\nonumber\\
&&\qquad 2\Re  {\it \bf e}  \{\vec{E}_{TX}(t,L)\cdot\vec{E}_{RB}^*(t,L)\}\Bigl]
\label{eq:i_out_00}
\end{eqnarray}
where $\vec{E}_{RB}(t,L) = {\Gamma} \sum_{k=1}^{k_{max}}\rho_k\vec{E}_{TX}(t,2 z_k+L)$, and where $\mathcal{R}$ is the Responsivity of the detector.

At the last side of Equation \eqref{eq:i_out_00}, the term $|E_{RB}(t,L)|^2$ results to be much smaller with respect to the others and will be neglected in the following. The detected current can therefore be approximated as:

\begin{align}\label{eq:i_out}
&i_{out}(t) \simeq \nonumber\\
&=\mathcal{R}\left[\left|\vec{E}_{TX}(t,L)\right|^2 + 2\Re  {\it \bf e}  \{\vec{E}_{TX}(t,L)\vec{E}_{RB}^*(t,L)\}\right]=&\nonumber\\
&=i_{out,TX}(t) + i_{out,TX,RB}(t)&
\end{align}

The first term at the right-hand side of Equation \eqref{eq:i_out} would coincide with the total detected current if the RB effect were absent, and is given by:

\begin{align}\label{eq:i_out_TX}
i_{out_{TX}}(t) = i_{out,0}[1+s(t-\tau_L)]e^{-\alpha L}
\end{align}

where {$i_{out,0}=\mathcal{R}|E_0|^2$} and  $\tau_L = \hat{\tau} L$. The second term at the right-hand side of Equation \eqref{eq:i_out} is instead the one responsible of the presence of the RB-induced spurious terms in the current power spectrum. From Equations \eqref{eq:campo_no_scattering} and \eqref{eq_E_RB}, its expression is given by:
{\begin{align}\label{eq:i_out_TX_RB_no_approx}
&i_{out_{TX,RB}}(t)= e^{-\alpha L} {\Gamma}i_{out,0} \times \nonumber\\
&2\Re  {\it \bf e}  \left\{\sum_{k=1}^{k_{max}}\rho^*_k\sqrt{1 + s(t-\tau_L)}\sqrt{1+s(t-2\tau_k-\tau_L))}  \times \right.\nonumber\\
&\left.e^{j[\theta(t-\tau_L)-\theta(t-2\tau_k-\tau_L)+2\beta z_k+\Delta\phi(t,2\tau_k)]-\alpha z_k}\right\}
\end{align}}
where $\tau_k = \hat{\tau} z_k$, $\Delta\phi(t,2\tau_k)=\phi(t-\tau_L)-\phi(t-\tau_L-2\tau_k)$, and where, from Prosthaphaeresis Formulas, it can be derived (see Equation \eqref{eq:theta}) $\theta(t-\tau_L)-\theta(t-2\tau_k-\tau_L)=\sum_{i=1}^{N_{s}}2M(\omega_{RF,i})\sin{\left(\omega_{RF,i}\tau_k\right)} \cos{\left(\omega_{RF,i}(t-\tau_L-\tau_k)\right)}$.

{Note that, depending on the expression of $s(t)$, and in particular on the applicability of the assumption $OMI_i<<1,\forall\, i$ (see Equations \eqref{eq:I_RF} and \eqref{eq:N_p}), the prevailing contribution to the nonlinearities can be ascribed either to $i_{out_{TX,RB}}(t)$ or to $i_{out_{TX}}(t)$.}

{Indeed, if $OMI_i<<1,\forall\, i$, (as for the case e.g. of SKA-LOW, where the optical modulation index is generally below 1\%), $i_{out_{TX}}(t)$ can be assumed as a linear function of the modulating signal, since the terms related to the nonlinearity of the photon-current laser curve, which are proportional to $(i_{RF}(t))^2$, $(i_{RF}(t))^3, \dots$ can be neglected.}


{Exploiting such relationship, approximations can be performed at the second side of Equation \eqref{eq:i_out_TX_RB_no_approx}, where both radicands can be just reduced to the unit number.} Indeed it can be straightforwardly derived that this approximation does not prevent to put into evidence the prevailing contributions to the distortion terms associated with the phenomenon under study. The quantity $i_{out_{TX,RB}}(t)$ assumes then the form:

{\begin{align}\label{eq:i_out_TX_RB_prima}
&i_{out_{TX,RB}}(t)\simeq e^{-\alpha L}  2{\Gamma}i_{out,0}\Re {\bf e}\left\{\sum_{k=1}^{k_{max}}\rho^*_k 
e^{j2\beta z_k-\alpha z_k} \times
\right.\nonumber\\
&\left. e^{j\sum_{i=1}^{N_{s}} x_{i,k} \cos{\left(\omega_{RF,i}(t-\tau_L-\tau_k)\right)}} e^{j\Delta\phi(t,2\tau_k)}\right\}
\end{align}}

 \noindent where $x_{i,k} =2M(\omega_{RF,i})\sin{\left(\omega_{RF,i}\tau_k\right)}$.
 
 {Applying now the Jacobi-Anger expansion $e^{ju\cos(\psi)}=\sum_{n=-\infty}^{\infty}(j)^n J_{n}(u)e^{jn\psi}$}, where $u=x_{i,k}$ and $\psi=\omega_{RF,i}(t-\tau_L-\tau_k)$, the expression of $i_{out_{TX,RB}}(t)$ finally becomes: 

{
\begin{align}\label{eq:i_out_TX_RB}
&i_{out_{TX,RB}}(t)\simeq e^{-\alpha L} 2{\Gamma} i_{out,0}\Re{\bf e}\left\{\sum_{k=1}^{k_{max}}\rho^*_k e^{j2\beta z_k-\alpha z_k}\times\right.\nonumber\\
&\left.
  \sum_{\substack{n_1,\dots,n_{N_s}=\\=-\infty}}^{+\infty}
{\left( j \right)^{n_1}J_{n_1}(x_{1,k})\dots \left(j\right)^{n_s}J_{n_{N_s}}(x_{N_s,k})} \times
\right.  \nonumber \\
& \left.\times e^{j\, (n_1\omega_{RF,1}+\dots+n_{N_s}\omega_{RF,N_s})(t-\tau_L-\tau_k)}e^{j\Delta\phi(t,2\tau_k)}\right\}
\end{align}}

{\noindent If, on the contrary the condition $OMI_i<<1,\forall\, i$ is not respected, then the approximations just performed cannot be applied. In this case the component $i_{out_{TX}}(t)$ can be considered as the main source of nonlinearity of the link, because of the presence of the terms of $s(t)$ multiplied by $a_2$, $a_3$, etc. Due to their clear identifiability, in the following paragraph the two different situations will be separately analyzed.}

\subsection{Determination of the output Power Spectrum}
{To analyze the impact of RB on signal distortion we proceed first in taking the square module of the Fourier Transform (FT) of $i_{out}$ (see Equation \eqref{eq:i_out}). Then, in order to obtain its Power Spectral Density (or {\it Power Spectrum}) $PSD_{out}$, the operator $\lim_{\cal{T}\rightarrow\infty}\frac{1}{\cal{T}}(\cdot)$ is applied, where $\cal{T}$ represents the integration interval of the Fourier Transform. This is necessary since the considered signal has not finite energy and therefore integration in the time domain diverges. Moreover, because of the presence of the statistical quantities $\rho_k$, the expected value has to be considered. Formally, the expression of $PSD_{out}$ can be written as:}

{
\begin{align}\label{eq:PSD}
&PSD_{out}=\lim_{\cal{T}\rightarrow\infty}\frac{1}{{\cal{T}}}\mathbb{E}\left[\left|\mathcal{F}_{\cal{T}}\{i_{out}\}\right|^2\right]=\nonumber\\
&=\lim_{\cal{T}\rightarrow\infty}\frac{1}{\cal{T}}\mathbb{E}\left[\left|\mathcal{F}_{\cal{T}}\left\{i_{out,TX}\right\}+\mathcal{F}_{\cal{T}}\left\{i_{out,TX,RB}\right\}\right|^2\right]\nonumber\\
&=\lim_{\cal{T}\rightarrow\infty}\frac{1}{\cal{T}}\left\{\mathbb{E}\left[\left|\mathcal{F}_{\cal{T}}\left\{i_{out,TX}\right\}\right|^2\right] + \mathbb{E}\left[\left|\mathcal{F}_{\cal{T}}\left\{i_{out,TX,RB}\right\}\right|^2\right]+ \right.\nonumber\\
&+\left. \mathbb{E}\left[2\Re  {\it \bf e}  \left\{\mathcal{F}_{\cal{T}}\left\{i_{out,TX}\right\}\mathcal{F}_{\cal{T}}^*\left\{i_{out,TX,RB}\right\}\right\}\right] \,\, \right\}
\end{align}
}

\noindent {where $\mathcal{F}_{\cal{T}}\{\cdot\}=\int_{-{\cal{T}}/2}^{{\cal{T}}/2} \, (\cdot) \, e^{-j \omega t}dt$ represents the FT operator computed in the integration interval $\cal{T}$}, and where, for simplicity, the explicit dependence on time of the currents to be transformed has been omitted.\\
\indent The last side of Equation \eqref{eq:PSD} is composed of three elements. {As a first consideration, the third term at the last side of Equation \eqref{eq:PSD} can be shown to be null, {exploiting the statistical properties of $\rho_k$ for which it is true that $\mathbb{E}[\rho_k] = 0\; \forall k$}. This reduces $PSD_{out}$ to the sum of the individual power spectra of $i_{out_{TX}}(t)$ and $i_{out_{TX,RB}}(t)$, named in the following $PSD_{TX}$ and  $PSD_{TX,RB}$, respectively.}

{Starting from $PSD_{TX}$, two different situations can be obtained, depending on the condition $OMI_i<<1,\forall\,i$. If this hypothesis is fulfilled, then the following expression of $PSD_{TX}$ can be written:}

\begin{align}
&PSD_{TX} (\omega)= {\lim_{\cal{T}\rightarrow\infty} \frac{1}{\cal{T}}}\left[e^{-2\alpha L}i_{out,0}^2 \left\{\delta(\omega) + 
 \sum_{i=1}^{N_{s}} \left(\frac{OMI_i}{2}\right)^2 \right.\right. \nonumber \\
&\left.\left. \times \left[\delta(\omega-\omega_{RF,i})+\delta(\omega+\omega_{RF,i})\right]\right\}\right]
\label{eq:PSD_TX}
\end{align}

where $\delta(\cdot)$ represents the Dirac generalized function (or {\it Dirac distribution}). {As mentioned above, $PSD_{TX}$ in this case is assumed to be ideal, meaning that the nonlinear contributions given by the terms $a_2$ and $a_3$ of equation \eqref{eq:N_p} are negligible.\\
\indent On the other hand, if the condition $OMI_i<<1,\forall\,i$ is not fulfilled, the expression of $PSD_{TX}$  will include all spurious terms which are linear combination of the frequencies employed. The expressions of some of such terms of particular interest will be given in the following subsection.}

The second term {in Eq. \eqref{eq:PSD}} represents the Power Spectral Density $PSD_{TX,RB}$ of the spurious terms generated by the RB feedback. Exploiting the general relation $\mathcal{F}\{\Re{\bf e}\{g(t)\}\} = \frac{1}{2}\left[\mathcal{F}\{g(t)\} + \mathcal{F}\{g^*(t)\} \right]$, and taking advantage of the statistical properties of $\rho(z)$ expressed by \eqref{eq:rho_k}, {which imply also $\mathbb{E}[\rho_k\rho_h] =0\quad\forall h,k$ }, after a lengthy but direct derivation, the expression of $ PSD_{TX,RB}$ can be determined as:

\begin{align}\label{eq:PSD_TX_RB}
& PSD_{TX,RB} (\omega)\simeq 2{\Gamma^2}i^2_{out,0}e^{-2\alpha L}\sum_{k=1}^{k_{max}}{\sigma^2_{\rho_k}} e^{-2\alpha z_k}\times\nonumber\\
&\left.
\times  \sum_{\substack{n_1,\dots,n_{N_s}=\\=-\infty}}^{+\infty}
{|J_{n_1}(x_{1,k})|^2\dots |J_{n_{N_s}}(x_{N_s,k})|^2} \times{\lim_{\cal{T}\rightarrow\infty}\frac{1}{\cal{T}}}
\right.  \nonumber\\
&\left[\delta\left(\omega-(n_1\omega_{RF,1}+\dots+n_{N_s}\omega_{RF,N_s})\right)*{\left|\mathcal{F}_{\cal{T}}\left\{e^{j\Delta\phi(t,2\tau_k)}\right\}\right|^2}\right]
\end{align}

where ($*$) is the convolution operator.
Exploiting the properties of the Dirac distribution, the last factor at the second side of Equation \eqref{eq:PSD_TX_RB} can be indicated as $\mathcal{L}(\omega-(n_1\omega_{RF,1}+\dots+n_{N_s}\omega_{RF,N_s}),2\tau_k)$, where {$\mathcal{L}(\omega,2\tau_k)=\lim_{\cal{T}\rightarrow\infty}\frac{1}{\cal{T}}\left|\mathcal{F}_{\cal{T}}\left\{e^{j\Delta\phi(t,2\tau_k)}\right\}\right|^2$} represents the equivalent linewidth of the optical field resulting from the interaction of $E_{TX}$ with the component of $E_{RB}$ which is delayed by $2\tau_k$.
Indicating with $\tau_{coh}$ the coherence time of the optical field, its expression can be determined \cite{Chraplyvy_phase_noise} as:

\begin{align}\label{eq:L_function}
    &{\mathcal{L} (\omega,2\tau_k)=}\nonumber\\
    &={e^{-\frac{2 \tau_k}{\tau_{coh}}} \left[\delta(\omega) +2\frac{\sin(\omega 2\tau_k)}{\omega}\right] + \frac{2\tau_{coh}}{1 + (\tau_{coh}\omega)^2}\left[1-e^{-\frac{2 \tau_k}{\tau_{coh}}}\right]}
\end{align}

where the value of $\tau_{coh}$ \cite{Saleh} can be determined from the numerical solution of the integral given by:

\begin{align}\label{eq:tcoh}
    \tau_{coh} = \int_{-\infty}^{\infty}{\frac{\left|<\vec{E}_{TX}(t)\cdot\vec{E}^*_{TX}(t-\xi)>\right|^2}{\left|<\vec{E}_{TX}(t)\cdot\vec{E}^*_{TX}(t)>\right|^2}d\xi}
\end{align}

Without direct modulation, this term represents the intrinsic coherence time of the laser source, mainly limited by the spontaneous emission. Under direct modulation, the spurious phase modulation due to frequency chirp {decreases the value of $\tau_{coh}$}, because of the spectrum broadening produced \cite{Chraplyvy}.\\
\indent The output power spectrum consists then in the sum of $PSD_{TX}(\omega)$ and $PSD_{TX,RB}(\omega)$, given respectively by Equations \eqref{eq:PSD_TX} and \eqref{eq:PSD_TX_RB}, {when the condition $OMI_i<<1\;\forall i$ is verified. If the aforementioned condition is not verified, then total power spectrum is considered composed only by $PSD_{TX}(\omega)$, in which, differently from Eq. \eqref{eq:PSD_TX}, the terms of second and third order are in this case included.} 


\subsection{Undesired spurious terms of particular interest}
\label{subsec:HDp_IMDq}
In many practical situations the number of RFIs is $N_s=1$ or $N_s=2$.\\
\indent In the first case, the power of the $p-th$ harmonic distortion term $HD_{p\omega_{RF,1}}$ caused by the nonlinear behavior{, considering separately the cases in which the condition $OMI_1<<1$ is respected or not, results respectively from the integral of equations \eqref{eq:PSD_TX} and \eqref{eq:PSD_TX_RB}  over a bandwith $B$ which tends to zero, centered in $p\omega_{RF,1}$. {The expression of $HD_{p\omega_{RF,1}}$} becomes}:

{\begin{eqnarray}
HD_{p\omega_{RF,1}}&&=2G_{AMP}R_L \times 2{\Gamma^2}i^2_{out,0}e^{-2\alpha L} \times\nonumber\\ 
&&  \times \sum_{k=1}^{k_{max}}\sigma^2_{\rho_k} e^{-2\alpha z_k}\times J^2_{p}(x_{1,k})\times\nonumber\\ 
&&\times\lim_{B \rightarrow 0}\smashoperator{\int\limits_{p\omega_{RF,1}-\frac{B}{2}}^{p\omega_{RF,1}+\frac{B}{2}}} \mathcal{L}(\omega-p\omega_{RF,1},2\tau_k)d\omega
\label{eq:HDp_TX_RB}
\end{eqnarray}}
{\noindent  if $OMI_{1,2}<<1$, while otherwise it becomes:}

\begin{flalign}
&{HD_{p\omega_{RF,1}} = G_{AMP}R_L\times i^2_{out,0}e^{-2\alpha L}a_p^2 \frac{OMI_1^{2p}}{2^{2p-1}}}
\label{eq:HDp_TX}
\end{flalign}
\noindent {where $R_L$ is the load resistance and $G_{AMP}$ is considered to take into account the possible presence of an amplifier right after the photodetector, as it is for the system considered in Section IV.}\\
\indent For the case {$OMI_1<<1$}, Eq. \eqref{eq:HDp_TX_RB} shows that $HD_{p\omega_{RF,1}}$ consists in the sum of the square of the $p-th$ order Bessel Functions of first kind $J_p$, appropriately evaluated and weighted over all the $k_{max}$ reflecting sections.
To give a quantitative idea, if one considers the undesired second harmonic ($HD_{2\omega_{RF,1}}$) generation of an RFI signal with angular frequency $\omega_{RF,1}=2 \pi \cdot 70 \, MHz$, possible typical values at this frequency of the related parameters are $K_f \simeq 220\,\frac{MHz}{mA}$, $I_{RF,1,0}=0.5 \, mA$. This leads to a value of $M(\omega_{RF,1})=2 \pi K_f(\omega_{RF,1})I_{RF,1,0}/\omega_{RF,1} \simeq 1.5$. It can be verified also graphically that for the various $k=1,\dots k_{max}$ (in correspondence to which the $\sin{\left(\omega_{RF,1}\tau_k\right)}$ randomly assumes values between -1 and 1) the values of $J^2_{2}(2\cdot1.5\cdot \sin{\left(\omega_{RF,1}\tau_k\right)})$ give in average a non negligible contribution to the final value of $HD_{2\omega_{RF,1}}$.\\
\indent{ Eq. \eqref{eq:HDp_TX} shows instead that if the condition $OMI_1<<1$ is not respected, $HD_{p\omega_{RF,1}}$ is proportional to the quantities $a_p^2$ and $OMI_1^{2p}$, and (see Eq.'s \eqref{eq:I_RF} and \eqref{eq:N_p}) for increasing amplitude of the modulating current $I_{RF,1,0}$ it grows, e.g., as $(I_{RF,1,0})^4$ if $p=2$ and as $(I_{RF,1,0})^6$ if $p=3$.}

Analogously, for $N_s=2$, the power of one {of} the possible  $q-th$ intermodulation distortion terms {$IMD_{u\omega_{RF,1}+v\omega_{RF,1}}$, where $q=|u|+|v|$ with $u,v=\pm 1, \dots, \pm (|q|-1)$}, results to be: 
{
\begin{align}
IMD&_{u\omega_{RF,1}+v\omega_{RF,2}}=\nonumber\\
&\quad=2G_{AMP}R_L \times 2\Gamma^2i^2_{out,0}e^{-2\alpha L}\times \nonumber\\
&\quad\times \sum_{k=1}^{k_{max}}\sigma^2_{\rho_k} e^{-2\alpha z_k} J^2_{u}(x_{1,k}) J^2_{v}(x_{2,k})\times \nonumber\\
&\quad\times 
\lim_{B \rightarrow 0}\smashoperator{\int  \limits_{u\omega_{RF,1}+v\omega_{RF,2}-\frac{B}{2}}^{u\omega_{RF,1}+v\omega_{RF,2}+\frac{B}{2}}} \mathcal{L}(\omega-u\omega_{RF,1}-v\omega_{RF,2},2\tau_k)d\omega
\label{eq:IMDq_TX_RB}
\end{align}}
{\noindent if $OMI_{1},OMI_{2}<<1$, while otherwise it results to be:}

{\begin{align}
IM&D_{u\omega_{RF,1}+v\omega_{RF,2}} = \nonumber\\
=&G_{AMP}R_L\times i^2_{out,0}e^{-2\alpha L}a_q^2OMI_{1}^{2|u|}OMI_{2}^{2|v|}\frac{q^2}{2^{2q-1}} \label{eq:IMDq_TX}
\end{align}}

\indent {Similarly to what has been observed with reference to the $HD_{p\omega_{RF,1}}$ terms,  Eq. \eqref{eq:IMDq_TX_RB} shows that for the case {$OMI_1<<1$}, $IMD_{u\omega_{RF,1}+v\omega_{RF,2}}$ consists in a weighted sum of products of the squares of the Bessel Functions of first kind of orders $u$ and $v$ ($J_u$ and $J_v$, respectively), for which the same quantitative considerations developed above can be applied.}

\indent{On the contrary, if the condition $OMI_1<<1$ is not respected, Eq. \eqref{eq:IMDq_TX} shows that $IMD_{u\omega_{RF,1}+v\omega_{RF,2}}$ is proportional to the quantities $a_q^2$, $OMI_1^{2|u|}$, $OMI_2^{2|v|}$, i.e. it is proportional to  $(I_{RF,1,0})^{2|u|}$ and to $(I_{RF,1,0})^{2|v|}$. }\\
\indent {Finally, note that in order to perform correctly the evaluation of equations \eqref{eq:HDp_TX_RB}, \eqref{eq:IMDq_TX_RB}, it is necessary to take adequately small values for the average distances $z_{k+1}-z_k$ between two consecutive reflection sections. Through various  simulation tests it has been verified that convergence to stable, reliable  results  can be achieved utilizing values of $10^{-4} m$ or less.}

\section{{Application contexts and operating conditions which maximize the impact of the RB-induced nonlinearities}}
\label{sec:Operating_Conditions}
From the considerations developed, it can be desumed that the level of a certain undesired spurious frequency term reaches important values when the corresponding value/values of the phase modulation index/indexes $M$ are such that the associated squared Bessel functions of first kind (see Eq.\eqref{eq:PSD_TX_RB}, or also Equations \eqref{eq:HDp_TX_RB}, \eqref{eq:IMDq_TX_RB}) exhibit relatively high values in correspondence to the argument given by $2M$.

\begin{figure}[hbtp]
\centering
\includegraphics[scale=0.3]{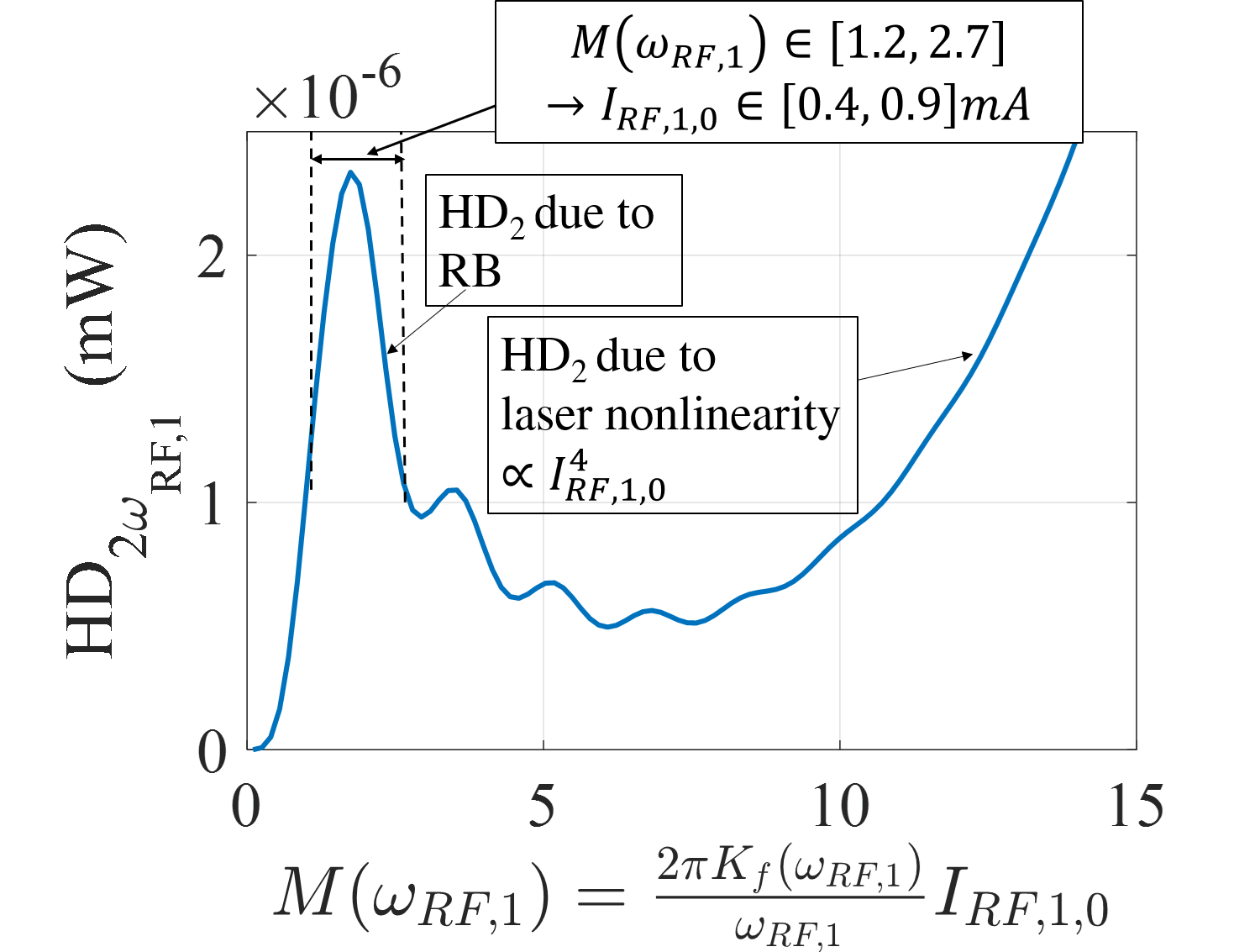}
\caption{Modelled values of $HD_{2\omega_{RF,1}}$ {(mW)} due to RB in a D-IMDD RoF link. The quantity in abscissa $M(\omega_{RF,1})=2 \pi \cdot K_f I_{RF,1,0}/\omega_{RF,1}$ is varied through $I_{RF, 1,0}$, keeping $K_f$ and $\omega_{RF,1}$ at constant values. For low values of $I_{RF, 1,0}$ $HD_{2\omega_{RF,1}}$ exhibits a behavior which can be related to $J_2^2(2M)$ (see Eq. \eqref{eq:HDp_TX_RB}), while for high values of  $I_{RF, 1,0}$ it results proportional to to $(I_{RF,1,0})^4$ (see Eq. \eqref{eq:HDp_TX}). }
\label{fig:operating_2d}
\end{figure}

To illustrate a practical example of this relationship, Figure \ref{fig:operating_2d} shows the computed behavior of $HD_{2\omega_{RF,1}}$ for $\omega_{RF,1}=2 \pi \cdot 70 \, MHz$, assuming $K_f=220 \, MHz/mA$, for varying values of $M_{\omega_{RF,1}}$. {The complete set of variables used for the simulations, which also refers to the system which will be employed in Section IV, is listed in Table \ref{tab:parameters}.}

{\begin{table}[t]
\centering
\caption{Simulation parameters.}
\renewcommand\arraystretch{1.2}{
\begin{tabular}{c|c|c}
\textbf{Symbol}& \textbf{Physical meaning}& \textbf{Value}\\
\hline
$I_{th}$        & Laser threshold current              & 9 $mA$\\
$I_{bias}$      & Laser bias current                   & 37 $mA$\\
$P_{opt}$       & Laser output power             & $6\,dB_m$\\
$Att_{iso}$          & Laser power isolation  & 20 $dB$\\
$\tau_{coh,0}$    & Laser unmodulated coherence time           & 63.6 $ns$\\
$G_L$           & Laser power amplification factor           & 18 $dB$\\
$K_{f}(70\, MHz)$  & Laser chirp factor at $70\, MHz$         & $220$ $MHz/mA$\\
$K_{f}(10\, KHz)$  & Laser chirp factor at $10 KHz\, MHz$         & $450$ $MHz/mA$\\
$a_2$           & Laser 2nd order non-linearity coeff. & $0.01$\\
$a_3$           & Laser 3rd order non-linearity coeff. & $0.09$\\
$\alpha$        & Fiber Power attenuation factor       & $10^{-4}$ $neper/m$\\
$\mathcal{S}$   & Fiber Recapture factor               & $10^{-3}$\\
$L$             & Fiber Length                   & 10 $Km$\\
$\mathcal{R}$   & PD Responsivity                & 1 $A/W$\\
$R_L$           & Load Resistance                & 50 $\Omega$\\
$G_{AMP}$       & Amplifier Power Gain                 & 22 $dB$\\
\hline
\end{tabular}}\label{tab:parameters}
\end{table}}

\begin{table*}[ht]
\centering
\caption{{Radioastronomic facilities formed by a high number of antennas, operating at frequencies of tens/hundreds of $MHz$, and proneness of their respective Antenna-to-Processing Room DownLinks (DL) to be affected by RB-induced nonlinearities}}
{
\renewcommand\arraystretch{1.2}{
\begin{tabular}{c|c|c|c|c|c}
\hline
Name & 
Location & 
Bandwidth & 
DL Length & 
DL Technology & 
Potentially affected \\ 
\hline
CHIME \cite{CHIME} & Canada & 400-800 $MHz$ & Up to 100 $m$ & RoF & No\\
GMRT \cite{GMRT} & India & 50-1500 $MHz$ & Up to 20 $Km$ & RF/IF over Fiber & Yes\\
HERA \cite{HERA} & South Africa & 100-200 $MHz$ & 150 $m$ / 500 $m$ & RF over Coax/RoF & No\\
LOFAR \cite{LOFAR} & 
Netherlands & 
10-250 $MHz$ & 
$\le 200$ m & 
RF over Coax & 
No\\
MWA \cite{MWA}& 
Australia & 
80-300 $MHz$ & 
$\le 5$ $km$ & 
RF over Coax / RoF & 
Yes\\
OVRO LWA \cite{OVRO_LWA} & 
USA & 
27-85 $MHz$ &
Up to few $km$ & 
RF over Coax / RoF & 
Yes\\
SKA-LOW \cite{SKA_LOW}& 
Australia & 
50-350 $MHz$ & 
$\le 10$ $km$ & 
RoF & 
Yes\\
\hline
\end{tabular}}}\label{tab:Radioastronomy}
\end{table*}

From the figure it can be appreciated that the $HD_{2\omega_{RF,1}}$ due to RB takes place in the range of values of $M$ which correspond approximately to $I_{RF,1,0} \in [0.4, 0.9]\,mA$, which in turn correspond to values  of input RF power levels $P_{RF,1,in}=\frac{1}{2}Z_{in}(I_{RF,1,0})^2$ (with $Z_{in}=50\,\Omega$ input impedance to the laser transmitter) ranging from $\sim -24\,dB_m$ to $ \sim -17\,dB_m$. 
Increasing $M$ (i.e. increasing $I_{RF,1,0}$) the value of $HD_{2\omega_{RF,1}}$ decreases, due to the fact that $J_2^2(2M)$ exhibits in this case lower values. {Proceeding further to greater values of   $M$ (i.e. of $I_{RF,1,0}$), it can be observed that starting from $M$ around $8$, which corresponds to $I_{RF,1,0} \sim 2.7 \,mA$ and to $P_{RF,1,in} \sim -7.4\,dB_m$,  $HD_{2\omega_{RF,1}}$ starts again to increase. This is due to the intrinsic nonlinearity of the laser source, and, as observed in Subsection \ref{subsec:HDp_IMDq} with regard to Eq. \eqref{eq:HDp_TX}, it features a behavior proportional to $(I_{RF,1,0})^4$.}

\begin{figure}[hbtp]
\centering
\includegraphics[scale=0.31]{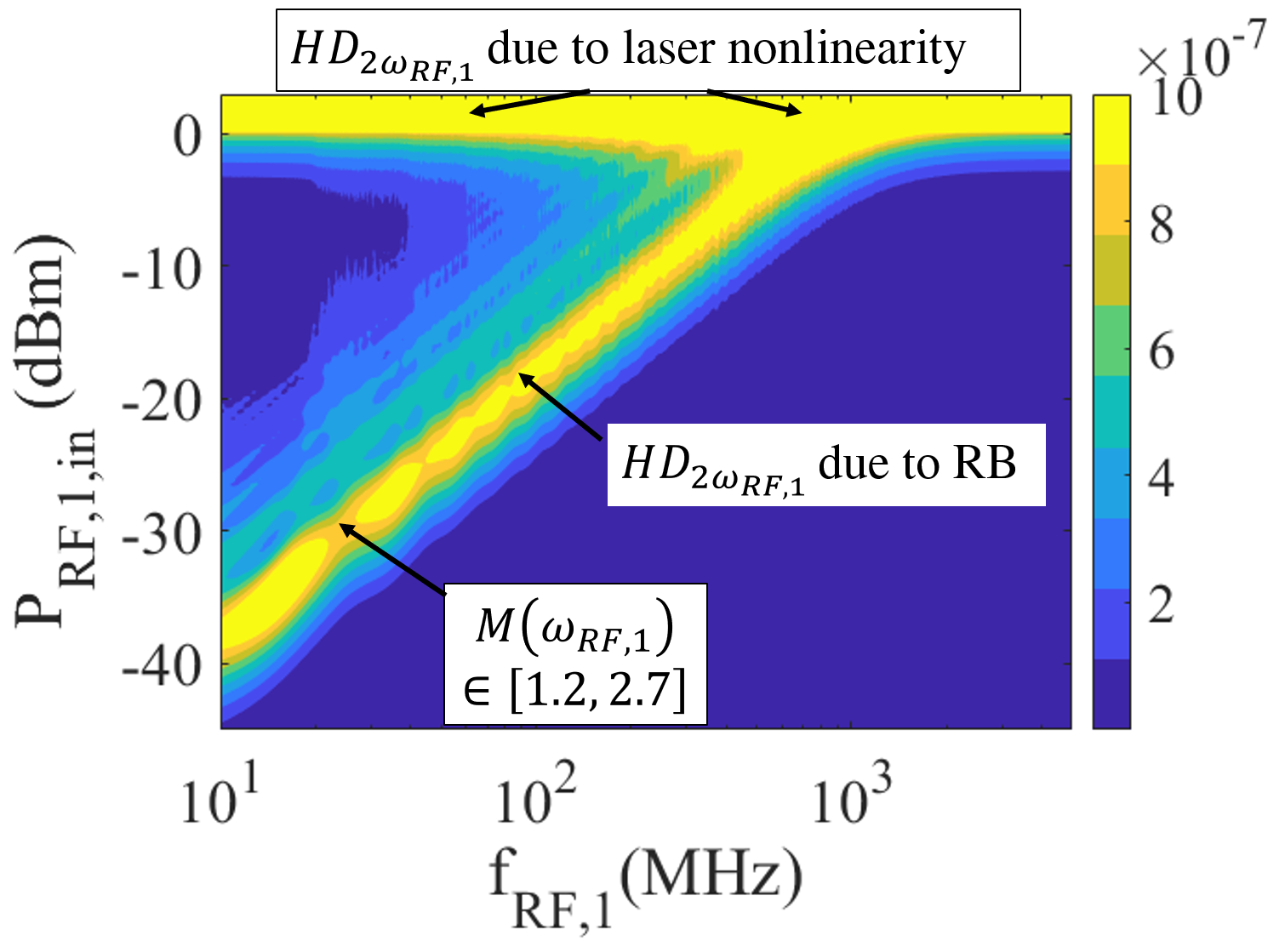}
\caption{Modelled values of $HD_{2\omega_{RF,1}}$ ($mW$) in the D-IMDD RoF link, whose parameters are reported in Table \ref{tab:parameters}, for varying values of both $f_{RF,1}=\omega_{RF,1}/(2 \pi)$ and $P_{RF, 1,in}$, keeping $K_f$ at constant values.}
\label{fig:operating_3d}
\end{figure}

{Focalizing again the attention on the RB-induced nonlinearities of which Fig. \ref{fig:operating_2d} visualizes an example, it can be observed that, generally speaking, the phenomenon studied can be found in any RoF system. 
However, the simultaneous presence of diverse aspects, which are listed below, must be met, in order to encounter the problem described.
\begin{enumerate}
    \item 	{\it Single downlink cost limitations} (e.g. in the order of $100\$$ or less for a single front-end receiver). 
    This constraint implies that no external modulators are utilized nor optical isolators are inserted in addition to the one embedded in the laser. The first  deficiency results in the presence of frequency chirp due to direct laser modulation, while the second implies that a portion of the sent signal is able re-enter the laser after Rayleigh back scattering. 
	\item {\it Relatively low power of the RF tones at the input section of the lasers.}  In the representative example visualized in Fig. \ref{fig:operating_2d}, values of $P_{RF,1,in} \sim -20\,dB_m$ or less allow to appreciate the phenomenon. Powers of these orders  of magnitude are low if compared with the ones used for e.g. typical RoF systems designed for telecommunications applications.  In this last case, the higher RF input powers (e.g. around $0\,dB_m$) would indeed "mask" the phenomenon considered,  even if the RoF system is not equipped with external modulators or additional isolators. 
\item {\it Relatively Low values of the RF frequencies transmitted in the optical channel} (from few MHz to hundreds of MHz).
	As recalled in the beginning of the present Subsection,  the quantity governing the behavior of the spurious frequencies for a given input tone $\omega_{RF,1}$ is $M=2 \pi \cdot K_f I_{RF,1,0}/\omega_{RF,1}=2 \pi \cdot K_f \sqrt{2P_{RF,1,in}/Z_{in}}/ f_{RF,1}$. 
Considering again the realistic example visualized in Fig. \ref{fig:operating_2d},  the $HD_{2\omega_{RF,1}}$ power can be regarded as  appreciable when the value of $M$ ranges roughly between 1.2 and 2.7. In case of powers $P_{RF,1,in}$ which can range  from around $-40\,dB_m$ to around  $-20\,dB_m$ and  taking into account that  it can be taken  $K_f \sim 200-300 MHz/mA$,  such values of $M$ are reached with RF frequencies of a few tens/few hundreds of MHz. These frequencies do not belong to the ones transmitted by RoF systems designed for 4G/5G mobile communications. In this last case, the values assigned to the RF carrier frequencies start from around $700 MHz$ and extend at least to some GHz. At these frequencies, in order to have values of $M$ ranging roughly between 1.2 and 2.7 it would be necessary to have powers $P_{RF,1,in}$ in the vicinity of $0\,dB_m$ or more, which would cause the nonlinearity due to the laser to mask the phenomenon, as specified in the previous point.
\end{enumerate}
}
{Figure \ref{fig:operating_3d} visualizes in logarithmic scale the considerations just developed, indicating the possible set of operating frequencies and RF input powers which can give rise to the phenomenon.}

{It can be appreciated in such figure the locus where the RB-related nonlinearities exhibit important values, given by the yellow stripe which starts from left-below and goes to right-above. Note that the extreme up-right values of $f_{RF,1}$ and $P_{RF, 1,in}$  identified by such stripe, given by the  vicinities of $f_{RF,1} \sim 700 \, MHz$ and $P_{RF, 1,in} \sim -5\,dB_m$ could represent a possible set of values utilized within LTE/5G applications. Nevertheless, apart from this singular case, the figure confirms that the values of  $f_{RF,1}$ and $P_{RF, 1,in}$ identified by the locus, in their  majority are typically not utilized in RoF systems for telecommunications applications. }\\
\indent{Moreover, still from  Figure \ref{fig:operating_3d} it can be clearly appreciated that a further increase in  $P_{RF, 1,in}$ from the value $P_{RF, 1,in} \sim -5\,dB_m$ determines a failure in the fulfillment of the condition $OMI <<1$. This causes $HD_{2\omega_{RF,1}}$ to increase for any value of $f_{RF,1}$, due to the intrinsic nonlinearity of the laser source. }\\
\indent{Note finally, that the same considerations just developed and illustrated with reference to $HD_{2\omega_{RF,1}}$ can be applied in the analysis of any other undesired  spurious term generated within the D-IMDD RoF link.}\\
\indent{The analysis performed within this Section allows then to affirm that the applicative scenario which can feature the combined presence of all the aspects listed above is the one of large radioastronomic facilities, composed by a high number of antennas, where the observed bandwidths include frequencies of few tens/few hundreds of MHz. }\\
\indent{To give an idea of the importance of these radioastronomic distributed antenna systems, Table \ref{tab:Radioastronomy} reports the most important ones, evidencing how some of them, in relation to the characteristics of the downlink (DL) which connects each Antenna to the Central Processing Room, the insurgence of RB-induced nonlinearities can be potentially experienced.}

\section{Proposed countermeasure: dithering tone}
\label{sec:dithering}
With the aim to counteract the insurgence of the nonlinearities originating from RB in D-IMDD RoF system,  the solution proposed consists in the direct modulation of the laser source through a further sinusoidal tone, which exhibits a lower frequency with respect to the one of the RFIs and, through the cited chirping effect, performs a so called {\it dithering} of the laser frequency.\\
\indent Note that this solution has already been proposed in the past, with the aim to reduce interferometric noise in optical fiber systems, which includes also noise induced by RB {\cite{Pepeljugoski,Lazaro}}. 
This is in line  with the fact that, as mentioned in the Introduction, the study of the detriments {due to RB} has been substantially focused on the added noise introduced by the phenomenon, and the countermeasures have consequently been directed to the solution of this problem.
Unlike that, the nonlinearities due to RB have not so far been put into evidence, and this proposal to use a dithering tone to reduce to acceptable levels the impact of the spurious terms generated, highlights a novel additional beneficial effect related to the application of this technique.\\
\indent The model derived in the previous Section can adequately describe the effect of the additional modulation, which is assumed to be performed with current $I_{dith}(t)=I_{dith,0}\cos(\omega_{dith}t)$.\\
\indent Without loss of generality, for the sake of clarity, as  exemplary operating condition it will be considered the case where two modulating tones are present, one constituted by a RFI, $I_{RF,1,0}\cos(\omega_{RF,1}t)$  and one by the dithering tone  $I_{dith,0}\cos(\omega_{dith}t)$.

{The undesired spurious terms to be analyzed are the $HD_{2\omega_{RF,1}}$ and $HD_{3\omega_{RF,1}}$ located at angular frequency $2\cdot \omega_{RF,1}$ and $3\cdot \omega_{RF,1}$, respectively. From Eq. \eqref{eq:HDp_TX_RB}, the expressions to be considered are then:} 

{\begin{align}\label{eq:HD2}
 HD_{2\omega_{RF,1}}&=  G_{AMP}R_L\times2\Gamma^2 2i^2_{out,0}e^{-2\alpha L}\times\nonumber\\
&\times\sum_{k=1}^{k_{max}}\sigma^2_{\rho_k} e^{-2\alpha z_k}\times J^2_{0}(x_{dith,k}) J^2_{2}(x_{1,k})\times\nonumber\\
&\times\lim_{B \rightarrow 0}\smashoperator{\int\limits_{2\omega_{RF,1}-\frac{B}{2}}^{2\omega_{RF,1}+\frac{B}{2}}} \mathcal{L}(\omega-2\omega_{RF,1},2\tau_k)d\omega
\end{align}}
{\begin{align}\label{eq:HD3}
HD_{3\omega_{RF,1}}&=  G_{AMP}R_L\times2\Gamma^2 2i^2_{out,0}e^{-2\alpha L}\times\nonumber\\
&\times\sum_{k=1}^{k_{max}}\sigma^2_{\rho_k} e^{-2\alpha z_k}\times J^2_{0}(x_{dith,k}) J^2_{3}(x_{1,k})\times\nonumber\\
&\times\lim_{B \rightarrow 0}\smashoperator{\int  \limits_{3\omega_{RF,1}-\frac{B}{2}}^{3\omega_{RF,1}+\frac{B}{2}}} \mathcal{L}(\omega-3\omega_{RF,1},2\tau_k)d\omega
\end{align}}

{\noindent for $OMI_1<<1$ and otherwise:}

{\begin{eqnarray}
&&HD_{2\omega_{RF,1}} =  G_{AMP}R_L i^2_{out,0}e^{-2\alpha L}a_2^2 \frac{OMI_1^{4}}{8}
\label{eq:HD2_TX}\\
&&HD_{3\omega_{RF,1}} =  G_{AMP}R_L i^2_{out,0}e^{-2\alpha L}a_3^2 \frac{OMI_1^{6}}{32}
\label{eq:HD3_TX}
\end{eqnarray}}

The beneficial effect of having introduced the dithering tone is related to the value assumed by the quantity $x_{dith,k}$ with respect to $x_{1,k}$. Indeed, taking e.g. the same  values chosen in Subsection \ref{subsec:HDp_IMDq} for  $\omega_{RF,1}$ and related quantities, the same average non negligible contribution to the final value of $HD_{2\omega_{RF,1}}$
is given by the terms $ J^2_{2}(2\cdot1.5\cdot \sin{\left(\omega_{RF,1}\tau_k\right)})|_{k=1,\dots,k_{max}}$.

On the contrary, the corresponding quantities related to $\omega_{dith}$ assume different values. In particular, being $\omega_{dith}\simeq 2 \pi \cdot 10 \, kHz$, $K_{f}(\omega_{dith})\simeq 450 \frac{MHz}{mA}$, $I_{dith,0}\simeq 20 \, \mu A$, this give rise to {$M (\omega_{dith}) = M_{dith} \simeq 900$}. For the various $k=1,\dots k_{max}$ the values of $J^2_{0}(2\cdot 900\cdot \sin{\left(\omega_{dith}\tau_k\right)})$ give in average an extremely low value, which in turn multiplies $J^2_{2}(2 \cdot M_{\omega_{RF,1}}\cdot \sin{\left(\omega_{RF,1}\tau_k\right)})$ and $J^2_{3}(2 \cdot M_{\omega_{RF,1}}\cdot \sin{\left(\omega_{RF,1}\tau_k\right)})$, reducing to negligible values the global contribution to $HD_{2\omega_{RF,1}}$ and $HD_{3\omega_{RF,1}}$.  

\begin{figure}[hbtp]
\centering
\begin{subfigure}{0.24\textwidth}
\includegraphics[width=\textwidth]{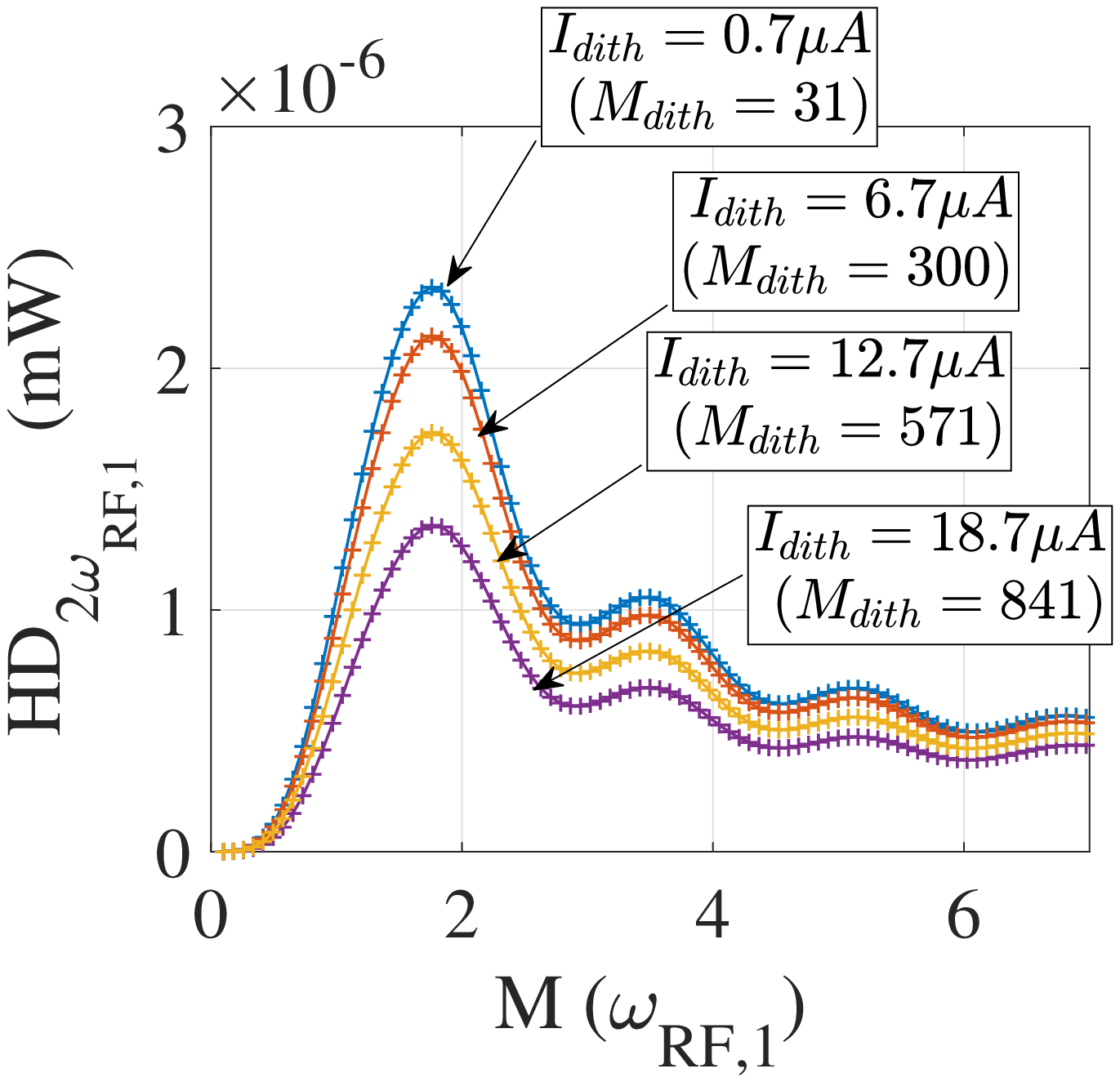}
\caption{}
\label{fig:HD2_MRF_varying_dithering_modello}
\end{subfigure}
\begin{subfigure}{0.24\textwidth}
\centering
\includegraphics[width=\textwidth]{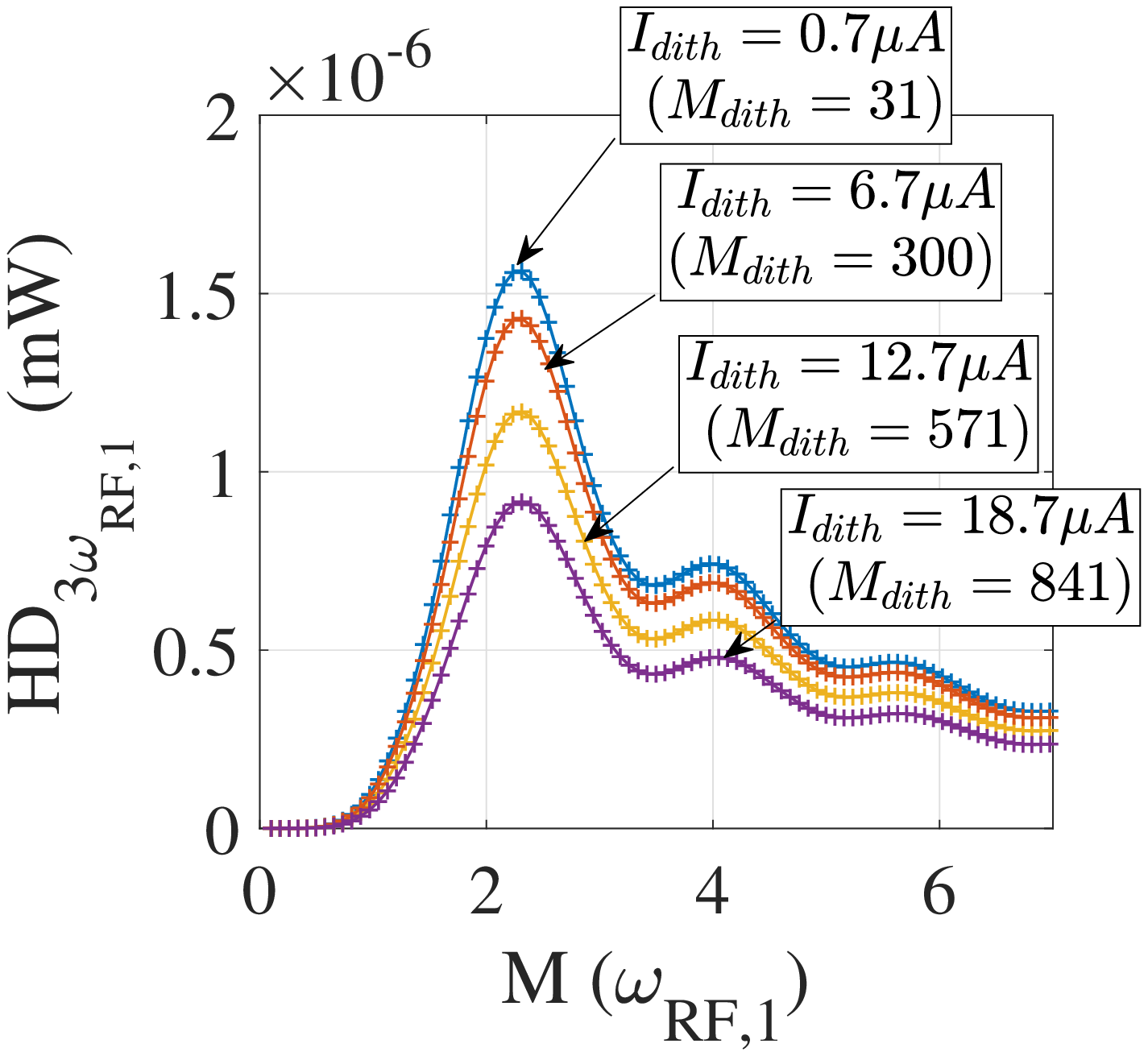}
\caption{}
\label{fig:HD3_MRF_varying_dithering_modello}
\end{subfigure}
\caption{Modelled values of $HD_{2\omega_{RF,1}}$ and $HD_{3\omega_{RF,1}}$ due to RB in a D-IMDD RoF link,  where the introduction of a dithering tone progressively reduces the impact of these undesired distortion terms. See text for details.}
\end{figure}

\begin{figure}[t]
\centering
\begin{subfigure}{0.42\textwidth}
\centering
\includegraphics[width=\textwidth]{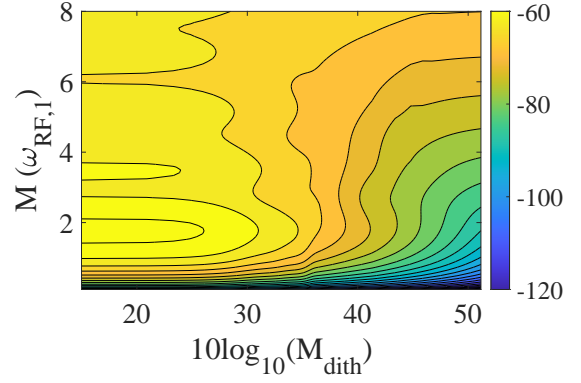}
\caption{}\label{fig:HD2_3D_simulations}
\end{subfigure}
\begin{subfigure}{0.42\textwidth}
\centering
\includegraphics[width=\textwidth]{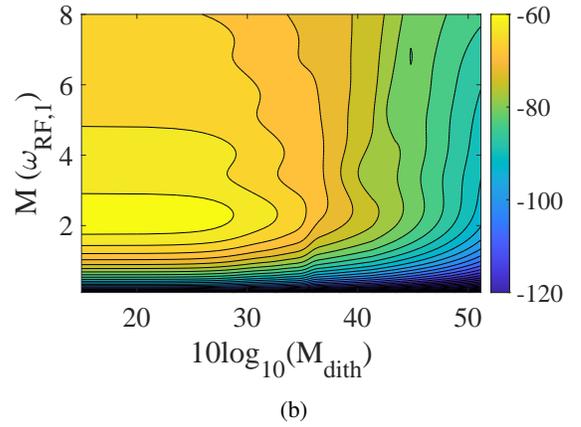}
\caption{}\label{fig:HD3_3D_simulations}
\end{subfigure}
\caption{Simulation of $HD_{2\omega_{RF,1}}$ (a) and $HD_{3\omega_{RF,1}}$ (b) expressed in $dB_m$ obtained varying both $M (\omega_{RF,1})$ and $M_{dith}$. See text for details.}
\end{figure}

Figures \ref{fig:HD2_MRF_varying_dithering_modello} and  \ref{fig:HD3_MRF_varying_dithering_modello} illustrate the theoretically computed behaviors of $HD_{2\omega_{RF,1}}$ and $HD_{3\omega_{RF,1}}$, respectively, showing their progressive reduction when the value of $I_{dith,0}$ is increased. The values of the parameters utilized are $\omega_{dith} = 2 \pi \cdot 10 \, kHz$, $K_{f}(\omega_{dith})= 450\, \frac{MHz}{mA}$,  $\omega_{RF,1} = 2 \pi \cdot 70 \, MHz$, $K_{f}(\omega_{RF,1})= 220\, \frac{MHz}{mA}$.  Behaviors refer to the same quantities are reported in Fig. \ref{fig:HD2_3D_simulations} and Fig. \ref{fig:HD3_3D_simulations} in 3D fashion. 

\begin{figure}
\centering
\includegraphics[scale=0.4]{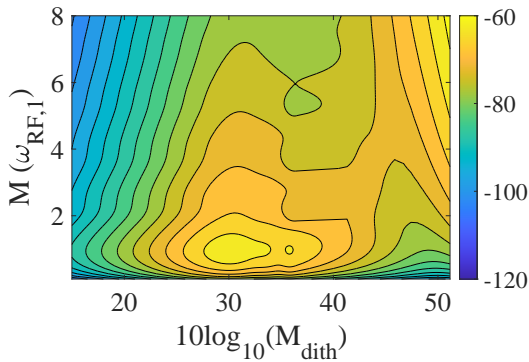}
\caption{Simulation of $IMD_{\omega_{RF,1}+\omega_{dith}}$ expressed in $dB_m$ obtained varying both $M (\omega_{RF,1})$ and $M_{dith}$.}
\label{fig:IMD_dith_3D_simulations}
\end{figure}

{It has however to be noted that, despite the great reduction obtained for $HD_{p\omega_{RF,1}}$, the insertion of the dithering tone is expected to generate an intermodulation itself with the RFI tone considered. \\
\indent This undesired effect must then be adequately kept under control. Its evaluation can be performed from equations \eqref{eq:IMDq_TX_RB} and \eqref{eq:IMDq_TX}, through which it is possible to evaluate, for example, the 2nd order intermodulation product between the dithering tone and the RFI tone, namely $IMD_{\omega_{RF,1}+\omega_{dith}}$, as follows: }
\begin{align}
IMD_{\omega_{RF,1}+\omega_{dith}} &=\nonumber\\
&=2 G_{AMP}R_L \times 2\Gamma^2i^2_{out,0}e^{-2\alpha L}\times\nonumber\\
&\times\sum_{k=1}^{k_{max}}\sigma^2_{\rho_k} e^{-2\alpha z_k} J^2_{1}(x_{1,k}) J^2_{1}(x_{dith,k})\times \nonumber\\
&\times 
\lim_{B \rightarrow 0}\smashoperator{\int  \limits_{\omega_{RF,1}+\omega_{dith}-\frac{B}{2}}^{\omega_{RF,1}+\omega_{dith}+\frac{B}{2}}} \mathcal{L}(\omega-\omega_{RF,1}-\omega_{dith},2\tau_k)d\omega
\label{eq:IMD_dith_TX_RB}
\end{align}
\noindent if $OMI_{1},OMI_{dith}<<1$, and otherwise:
\begin{equation}
IMD_{\omega_{RF,1} + \omega_{dith}} =  G_{AMP}\frac{R_L}{2}\times i^2_{out,0}a_q^2OMI_{1}^{2}OMI_{dith}^{2}. \label{eq:IMD_dith_TX}
\end{equation}
Figure \ref{fig:IMD_dith_3D_simulations} shows the simulation results for $IMD_{\omega_{RF1,1}+\omega_{dith}}$, which can be compared with the quantities $HD_{2\omega_{RF,1}}$ and $HD_{3\omega_{RF,1}}$ reported in Fig. \ref{fig:HD2_3D_simulations} and Fig. \ref{fig:HD3_3D_simulations}. 
{This comparison shows clearly that in addition to the beneficial effect of reducing  $HD_{2\omega_{RF,1}}$ and $HD_{3\omega_{RF,1}}$ a not appropriate level of $M_{dith}$ can cause an undesired high value of  $IMD_{\omega_{RF1,1}+\omega_{dith}}$. This happens e.g for $10\log_{10}M_{dith} \simeq 30\dots 35$ and $M(\omega_{RF,1}) \sim 1 \dots 2$.  In this example the optimum choice of the dithering amplitude is the one that gives $10\log_{10}(M_{dith})\in[40-50]$ considering $M_{RF,1} \le 2$. This region allows indeed to reduce of about 20-30dB the level of $HD_{2\omega_{RF,1}}$ and $HD_{3\omega_{RF,1}}$ keeping $IMD_{\omega_{RF1,1}+\omega_{dith}}<-80\,dB_m$.}\\
{\indent The comparison among figures  \ref{fig:HD2_3D_simulations} and \ref{fig:IMD_dith_3D_simulations} allows to appreciate also the region where the nonlinearity of the laser response starts to give its contribution. This impact is visible at the top of Fig. \ref{fig:HD2_3D_simulations} and at the right-top of Fig. \ref{fig:IMD_dith_3D_simulations}, where $HD_{2\omega_{RF,1}}$ and $IMD_{\omega_{RF1,1}+\omega_{dith}}$ start to increase proportionally with $M(\omega_{RF,1})$ and with both $M(\omega_{RF,1})$ and $M_{dith}$, respectively. In particular, while for $HD_{2\omega_{RF,1}}$ this contribution is due only by the quantity $OMI_1$ (see Eq. \eqref{eq:HD2_TX}), for $IMD_{\omega_{RF1,1}+\omega_{dith}}$ it depends on both $OMI_1$ and $OMI_{dith}$.}\\
\indent {Note finally that while this range of $M_{dith}$ is given for a specific case, optimum design parameters of a generic system can be extrapolated applying directly the equations \eqref{eq:HDp_TX_RB},\eqref{eq:HDp_TX}, \eqref{eq:IMDq_TX_RB},\eqref{eq:IMDq_TX}.}

\section{{Experimental validation of the model presented and of the countermeasure proposed}}\label{sec:experimental}

\begin{figure}[hbtp]
\centering
\includegraphics[scale=0.26]{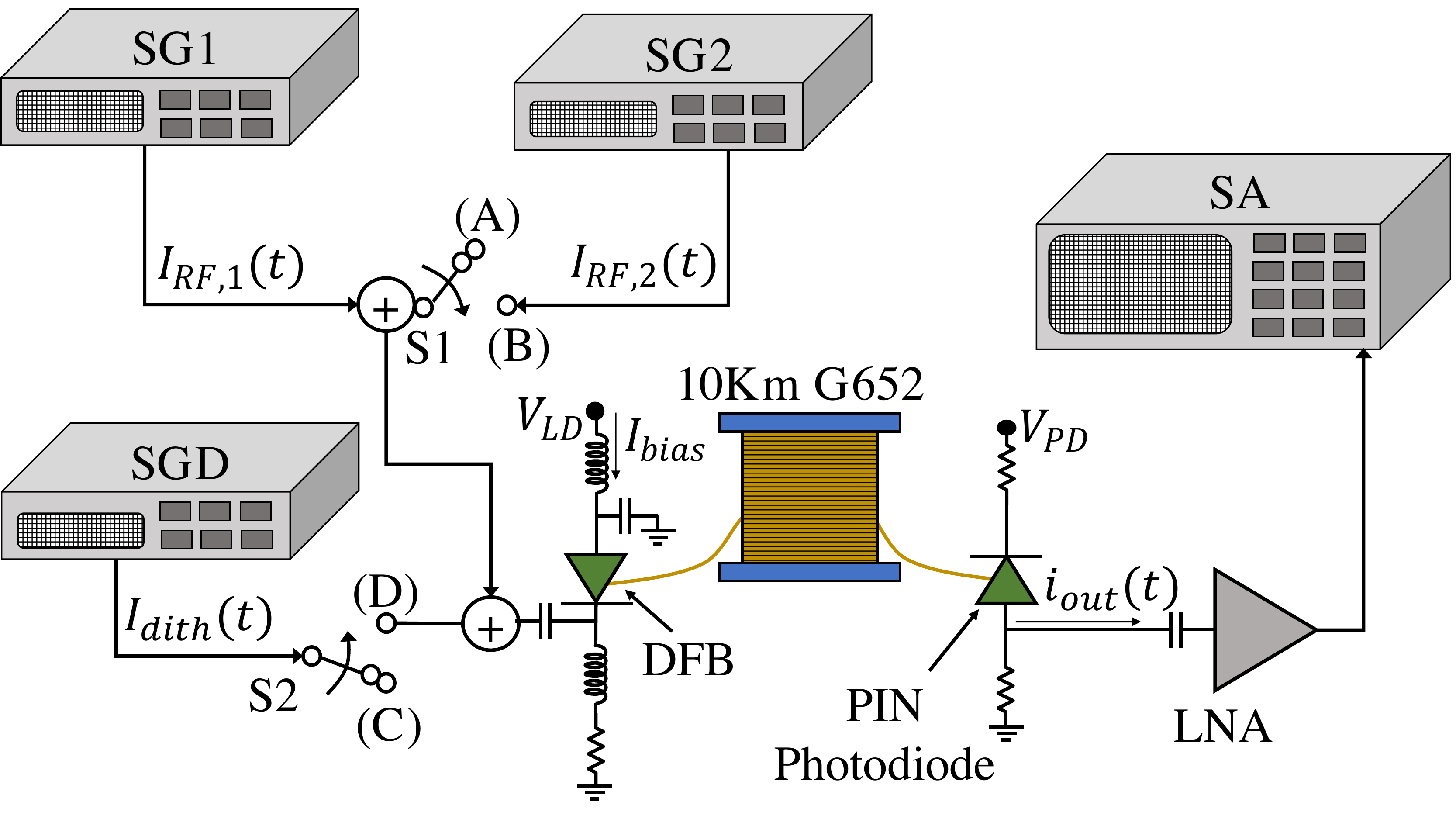}
\caption{Experimental setup utilized for the characterization of the nonlinearities due to RB in case of single tone and two tones with and without the use of the dithering tone. See text for details.}
\label{fig:setup}
\end{figure}

To analyze the impact of the nonlinearities produced by RB, the experimental setup shown in Figure \ref{fig:setup} has been utilized.

\begin{figure}[hbpt]
\centering
\begin{subfigure}{0.235\textwidth}
\centering
\includegraphics[width=\textwidth]{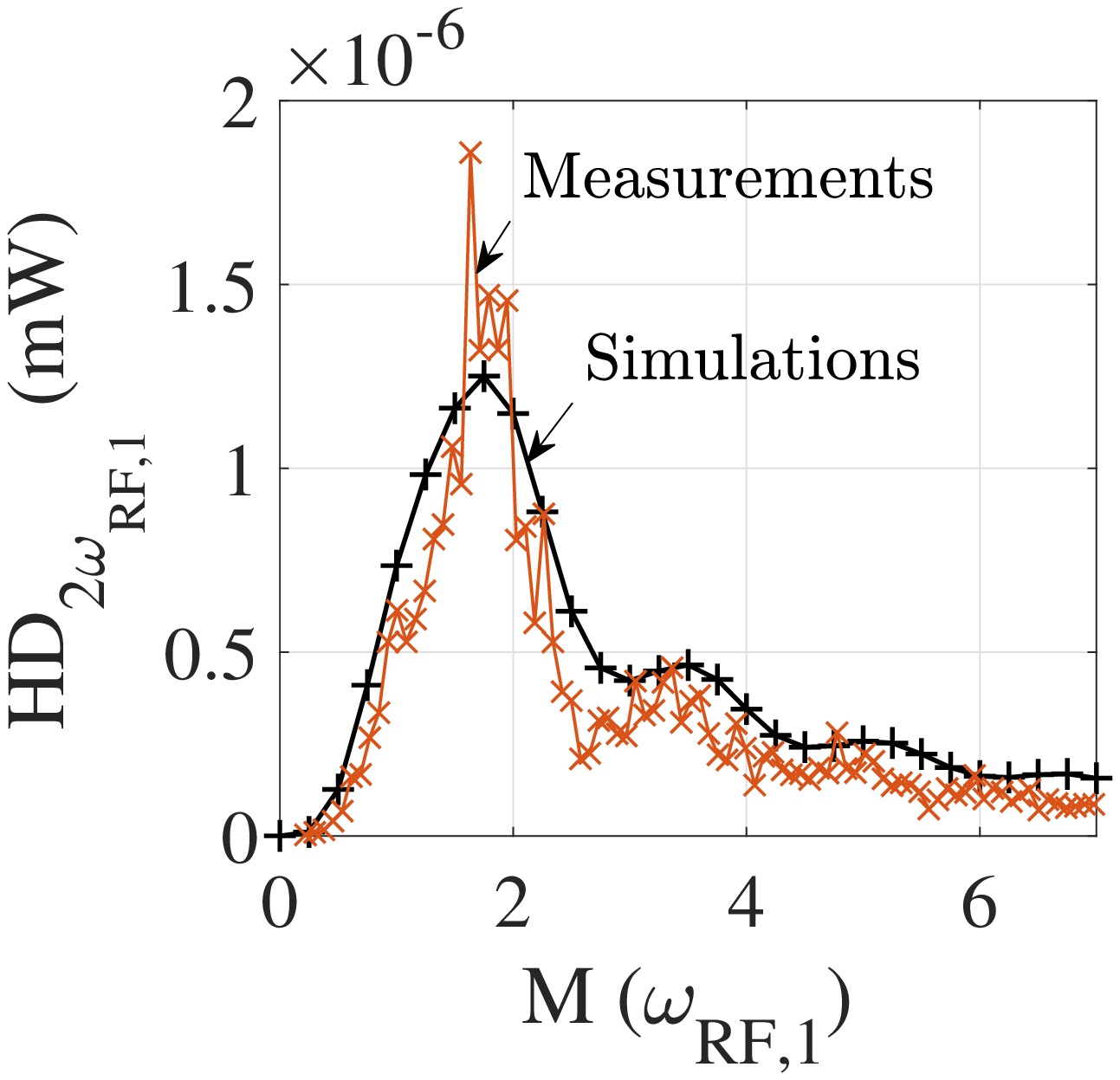}
\caption{}\label{fig:HD2_case1}
\end{subfigure}
\begin{subfigure}{0.235\textwidth}
\centering
\includegraphics[width=\textwidth]{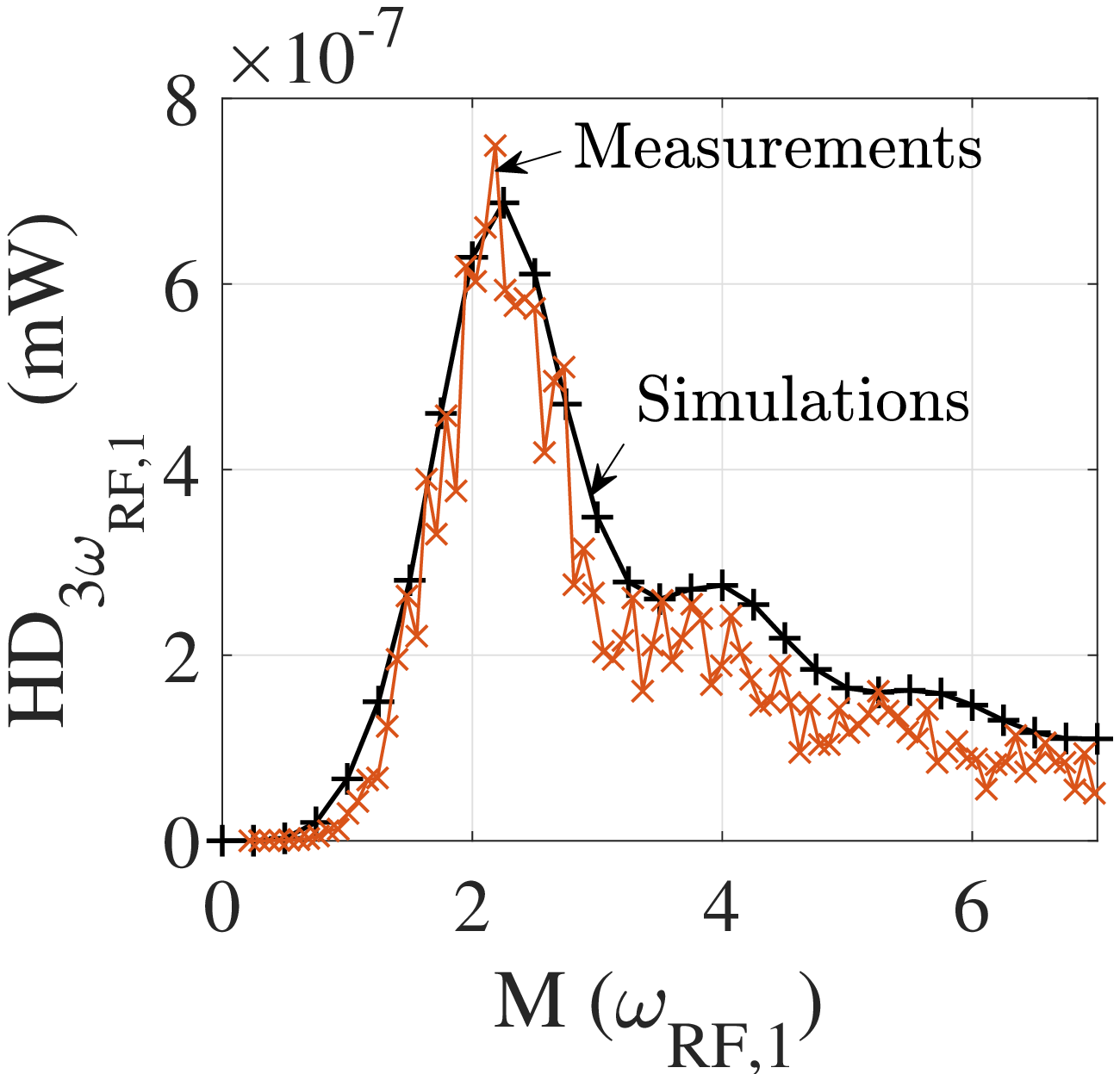}
\caption{}\label{fig:HD3_case1}
\end{subfigure}
\begin{subfigure}{0.235\textwidth}
\centering
\includegraphics[width=\textwidth]{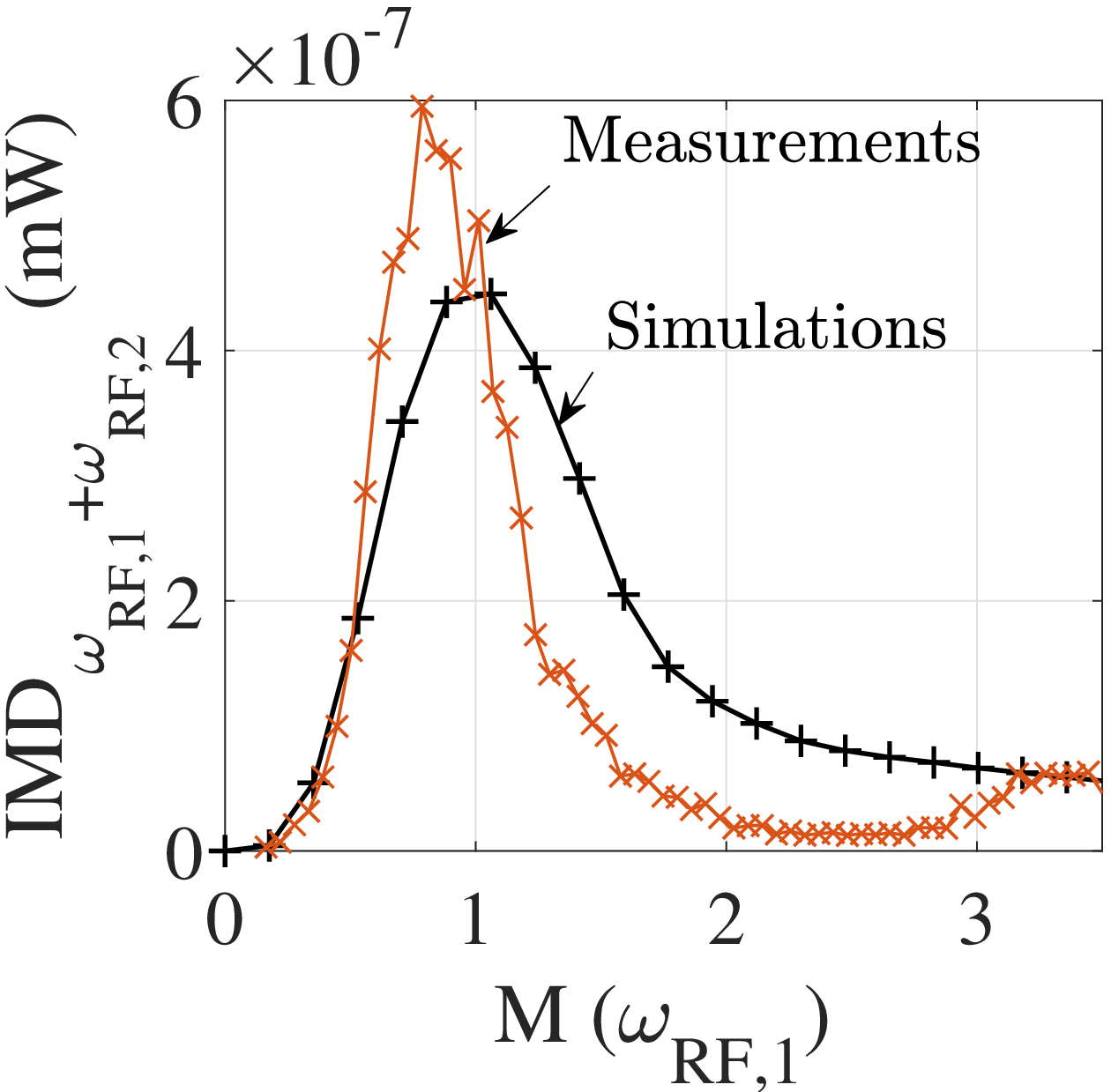}
\caption{}\label{fig:IMD2_case2}
\end{subfigure}
\begin{subfigure}{0.235\textwidth}
\centering
\includegraphics[width=\textwidth]{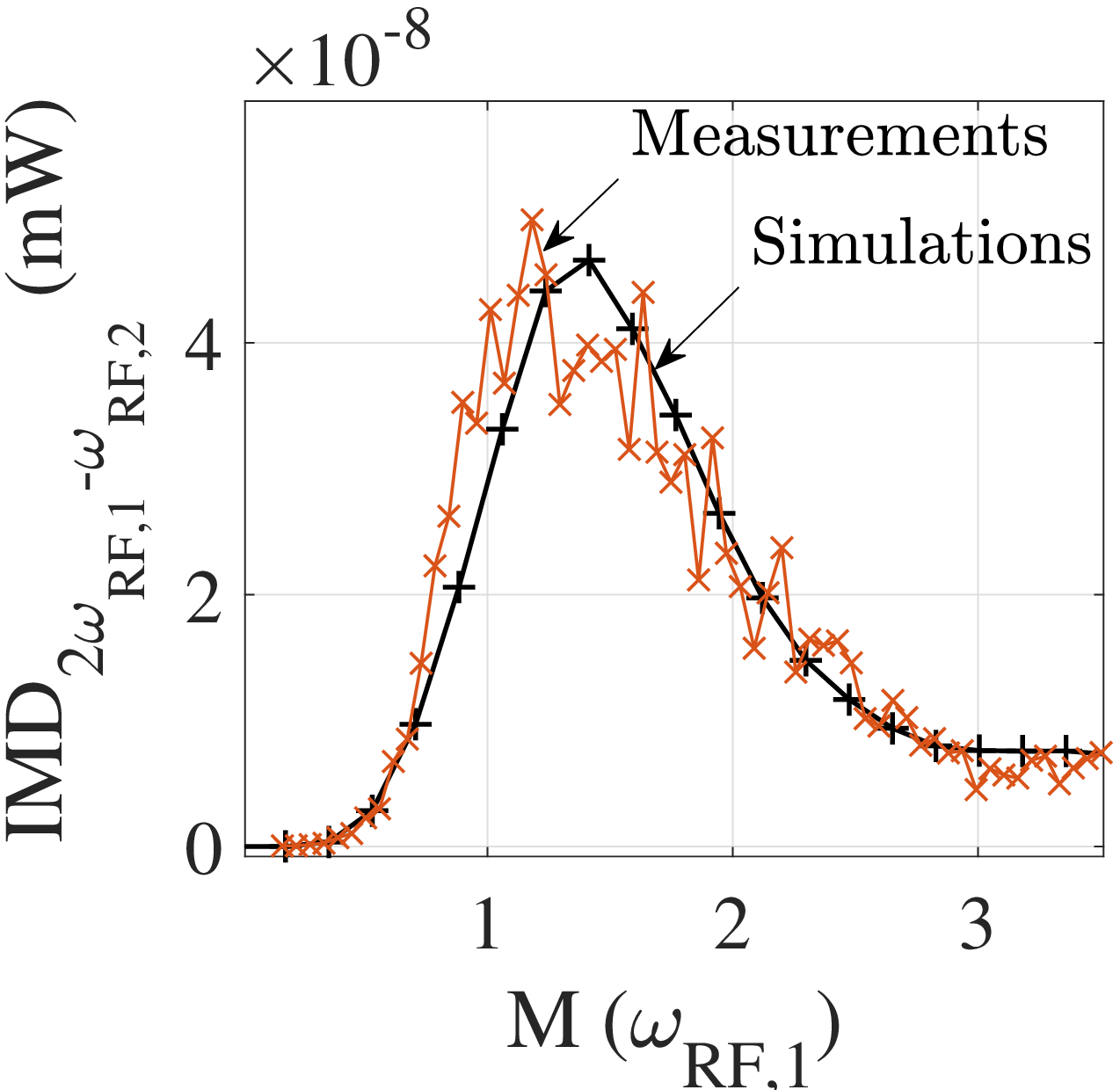}
\caption{}\label{fig:IMD3_case2}
\end{subfigure}
\caption{Comparison of measurements and simulation for $HD_{2\omega_{RF,1}}$ (a) and $HD_{3\omega_{RF,1}}$ (b) in presence of single tone modulation and $IMD_{\omega_{RF,1}+\omega_{RF,2}}$ (c) and $IMD_{2\omega_{RF,1}-\omega_{RF,2}}$ (d) in presence of two tones modulation.}
\end{figure}
\begin{figure}[hbpt]
\centering
\begin{subfigure}{0.241\textwidth}
\centering
\includegraphics[width=\textwidth]{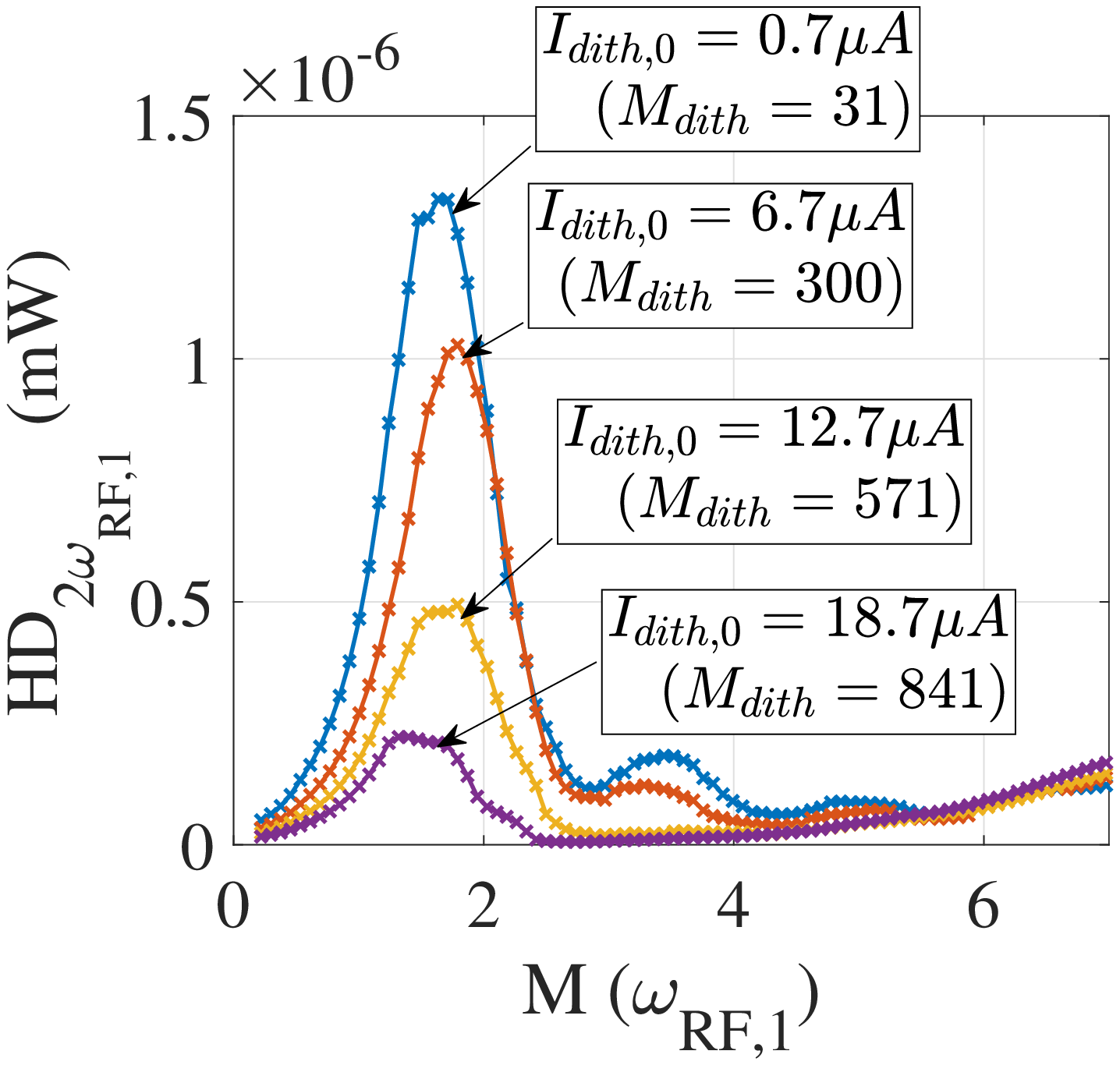}
\caption{}\label{fig:HD2_case3_meas}
\end{subfigure}
\begin{subfigure}{0.241\textwidth}
\centering
\includegraphics[width=\textwidth]{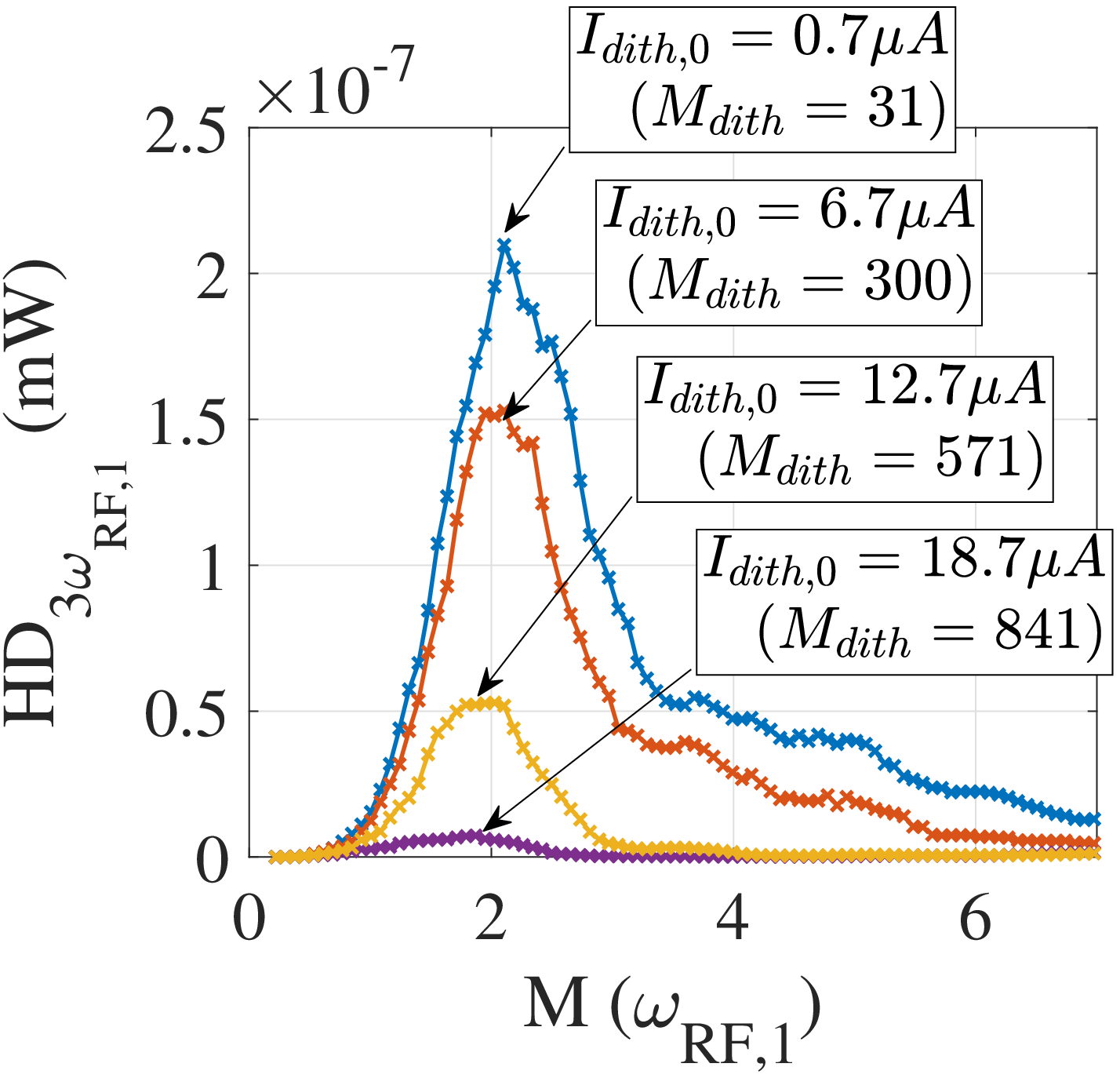}
\caption{}\label{fig:HD3_case3_meas}
\end{subfigure}
\caption{Impact of the dithering tone amplitude on the trend of $HD_{2\omega_{RF,1}}$ (a) and $HD_{3\omega_{RF,1}}$ (b) with respect to $M (\omega_{RF,1})$.}
\end{figure}
\noindent The optical link evaluated is composed of a 1310 $nm$ DFB source, operating at $I_{bias}-I_{th} = 28$ $mA$ with an optical output power of $6\,dB_m$, connected to a span of 10Km of G652 fiber followed by a PIN photodetector. These devices are directly connected each other by using APC connectors to minimize any possible further reflection in order to evaluate only the impact of RB.\\
\indent The input RF signal is generated through three signal generators SG1, SG2 and SGD which emit the signals $I_{RF,1}(t)$, $I_{RF,2}(t)$ and  $I_{dith}(t)$, respectively. The generators are connected to the switches S1, S2 and to RF couplers, inserted to analyze three different cases:

\vspace{0.2cm}
\begin{itemize}
    \item[(1)] \textit{Single tone modulation}:\hfill S1 $\rightarrow$ (A), S2 $\rightarrow$ (C).
    \item[(2)] \textit{Two tones modulation}: \hfill S1 $\rightarrow$ (B), S2 $\rightarrow $ (C).
    \item[(3)] \textit{Use of the dithering tone}: \hfill S1 $\rightarrow$ (A), S2 $\rightarrow$ (D).
\end{itemize}
\vspace{0.2cm}

\noindent Right after the PIN photodetector, a Low Noise Amplifier (LNA) with  $G_{AMP}= 22$ dB amplifies the RF component of $i_{out}$, while the final power spectrum of the signal coming out from the LNA is shown by a Spectrum Analyzer (SA).\\
\indent In order to perform a characterization with respect to the parameter $M$, the measurements have been performed by acting on the currents value variation, keeping fixed the frequencies of the tones. In case (1) the RF frequency was chosen to be $\omega_{RF,1}=2\pi\cdot 70$ $MHz$, in case (2) it was $\omega_{RF,1}=2\pi\cdot 65$ $MHz$ and $\omega_{RF,2} =2\pi\cdot 75$ $MHz$, and in case (3) it was  $\omega_{RF,1} =2\pi\cdot 70$ $MHz$ with $\omega_{dith} = 2\pi\cdot 10$ $kHz$. In all cases the value of the chirp factor for the chosen frequencies is $K_f(\omega_{RF,1})=K_f(\omega_{RF,2})=220$ $MHz/mA$,  while {for} the dithering frequency chosen {it is} $K_f(\omega_{dith})= 450$ $MHz/mA$. \\
\indent {The measurements of nonlinearities referring to cases (1) and (2) are shown in Figure \ref{fig:HD2_case1}, \ref{fig:HD3_case1} and \ref{fig:IMD2_case2}, \ref{fig:IMD3_case2}, respectively. In particular, for case (1) the measurements of $HD_{2\omega_{RF,1}}$ and $HD_{3\omega_{RF,1}}$ are shown, while for case (2) it is illustrated the behavior of $IMD_{\omega_{RF,1}+\omega_{RF,2}}$ and $IMD_{2\omega_{RF,1}-\omega_{RF,2}}$. In both cases the measurements are presented with respect to the quantity $M(\omega_{RF,1})$.}\\
{\indent The measured behaviors are compared with the correspondent simulated ones,  based on the mathematical model presented in Section \ref{sec:theoretical}. A good agreement can be in all cases appreciated between experimental and theoretical results.\\
\indent The effects of the implementation of the dithering tone are shown in Figures \ref{fig:HD2_case3_meas} and \ref{fig:HD3_case3_meas}. It can be observed that by increasing the dithering amplitude even of few $\mu A$s, a decrease for both $HD_{2\omega_{RF,1}}$ and $HD_{3\omega_{RF,1}}$, with respect to $M(\omega_{RF,1})$, is present, which is in agreement with the same trend described by the mathematical model shown in Section \ref{sec:theoretical}.} \\
{\indent Regarding the application under study, SKA-LOW, the specifications are given in terms of second and third order Output Intercept Points ($OIP_2$ and $OIP_3$ respectively), as it is typically done for RF systems, and in particular it must be $OIP_2>38\,dB_m$ and $OIP_3>28\,dB_m$ for the all downlink system (RF electronics and optical link). The quantities  $OIP_2$ and $OIP_3$ are directly related with $HD_{2\omega_{RF,1}}$ and  $HD_{3\omega_{RF,1}}$ by the following equations {\cite{Egan}:}
{
\begin{align}
&OIP_2 (dB_m) =\nonumber\\ &2\left[P_{out,1,0}(dB_m)\right]-\left[HD_{2\omega_{RF,1}}(dB_{m})+6(dB)\right]\label{eq:OIP2}\\
\nonumber\\
&OIP_3 (dB_m) =\nonumber\\ &\frac{3}{2}\left[P_{out,1,0}(dB_m)\right]-\frac{1}{2}\left[HD_{3\omega_{RF,1}}(dB_m)+9.54(dB)\right]\label{eq:OIP3}
\end{align}}
where in the present case it is possible to consider $P_{out,1,0}\simeq P_{RF,1,in}$ since the amplifier used compensates the losses of the optical link.\\
\indent Considering now values of input power that can range from $-50\,dB_m$ to $-20\,dB_m$ (i.e. $M(\omega_{RF,1})<2$), as for the RFI signals, it is possible to estimate what are the worst levels of $OIP_2$ and $OIP_3$ reached due to RB. In fact, note that, unlike the classical case studies of RF nonlinearities of 2-port devices, where the quantities $OIP_2$ and $OIP_3$ do not depend on the value of the input RF power, in presence of RB-induced nonlinearities this does not happen, and both $OIP_2$ and $OIP_3$ can significantly vary with the input RF power given.\\
\indent Figures \ref{fig:HD2_3D} and \ref{fig:HD3_3D} show the behavior of $HD_{2\omega_{RF,1}}$ (a) and  $HD_{3\omega_{RF,1}}$, respectively, by varying both $M(\omega_{RF,1})$ and $M_{dith}$, presenting the quantity in $dB_m$.
\begin{figure}[hbpt]
\centering
\begin{subfigure}{0.41\textwidth}
\centering
\includegraphics[width=\textwidth]{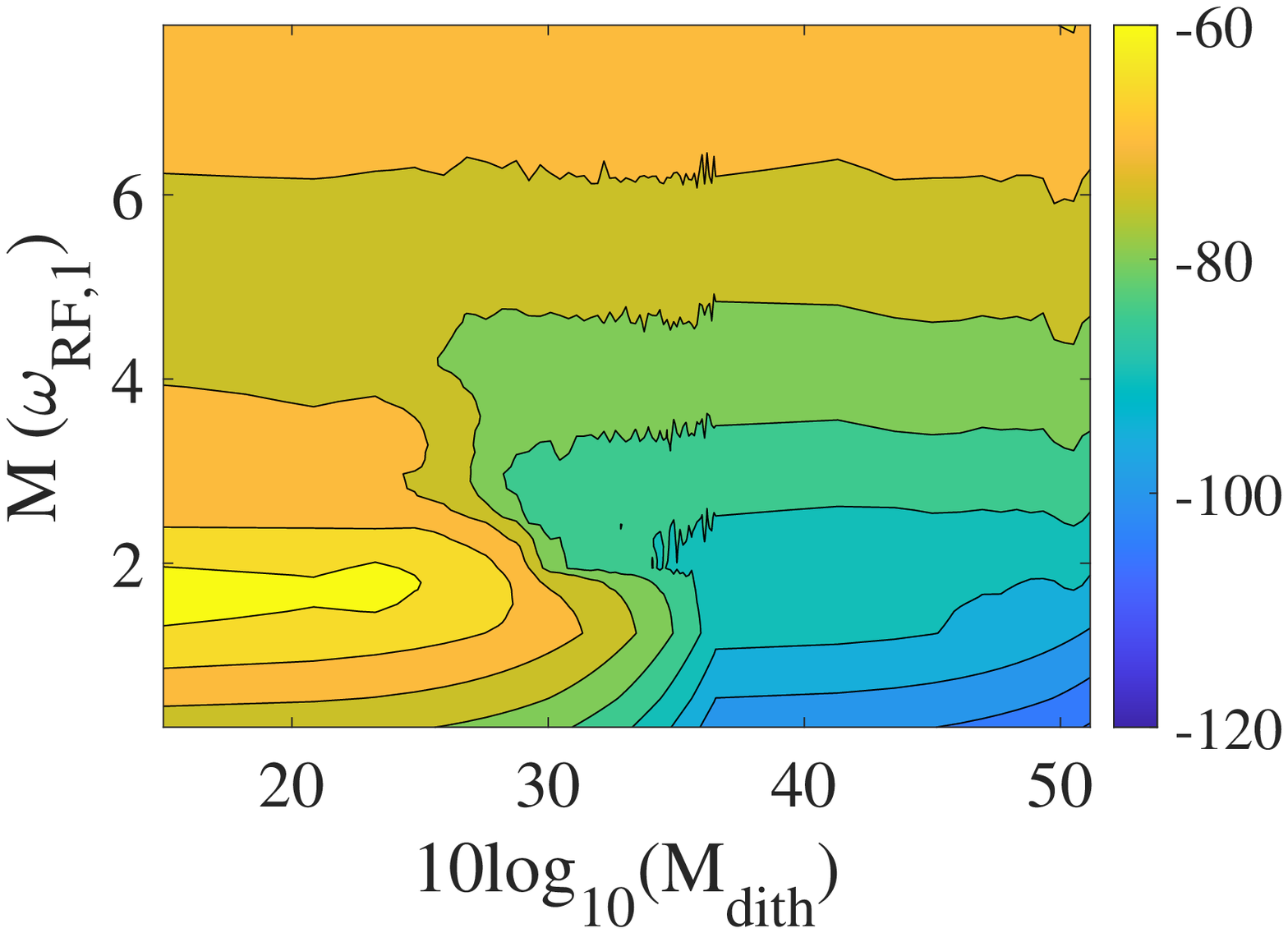}
\caption{}\label{fig:HD2_3D}
\end{subfigure}
\begin{subfigure}{0.41\textwidth}
\centering
\includegraphics[width=\textwidth]{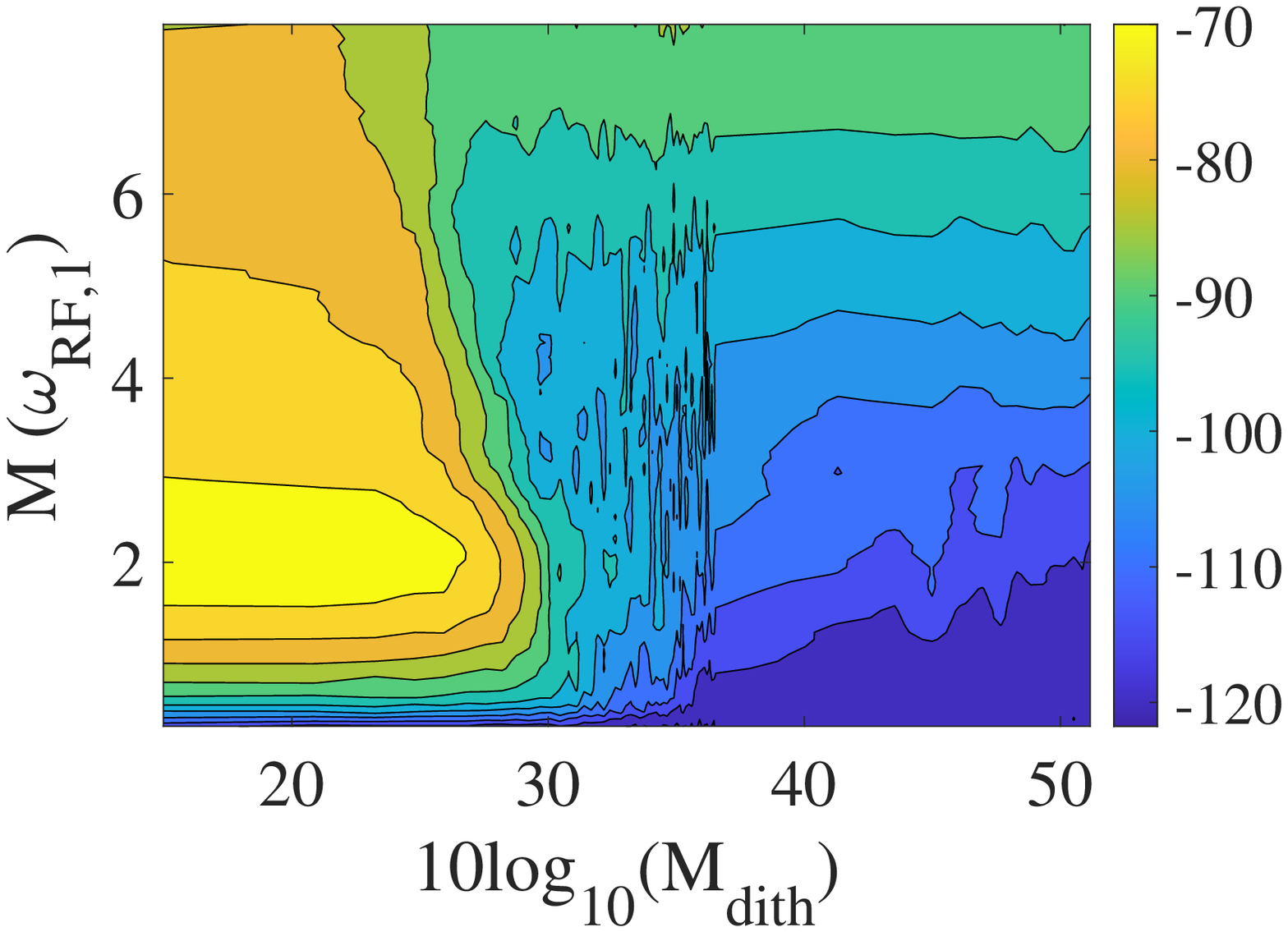}
\caption{}\label{fig:HD3_3D}
\end{subfigure}
\caption{Measurement of $HD_{2\omega_{RF,1}}$ (a) and  $HD_{3\omega_{RF,1}}$ (b) expressed in $dB_m$ obtained varying both $M (\omega_{RF,1})$ and $M_{dith}$.}
\end{figure}
\indent In particular, taking for example $M(\omega_{RF,1})\simeq 1$, which corresponds to $P_{in,1,0}=-26\,dB_m$, it is $HD_{2\omega_{RF,1}}\simeq -66\,dB_m$  and $HD_{3\omega_{RF,1}}\simeq -90\,dB_m$, which means $OIP_2\simeq 0\,dB_m$ and $OIP_3\simeq -5\,dB_m$ according to equations \eqref{eq:OIP2} and \eqref{eq:OIP3}.\\
\indent Indeed, these levels of $OIP_2$ and $OIP_3$ are far from being acceptable for the system and even in case it were possible for the electronic sections of the global receiver to be designed in order to satisfy the specifications of $OIP_2$ and $OIP_3$, this should be done at a high cost in terms of devices utilized and supply power absorbed.}
 
\begin{figure}[hbpt]
\centering
\includegraphics[scale = 0.42]{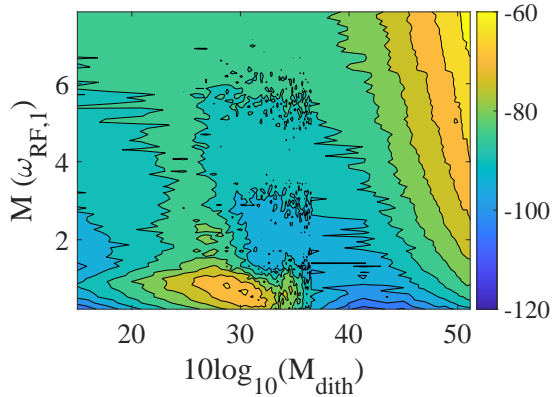}
\caption{Measurement of $IMD_{\omega_{RF,1}+\omega_{dith}}$ expressed in $dB_m$ obtained varying both $M (\omega_{RF,1})$ and $M_{dith}$.}\label{fig:IMD_dith_3D}
\end{figure}

{Figures \ref{fig:HD2_3D} and \ref{fig:HD2_3D} show instead that applying properly the dithering technique, it is possible to achieve $HD_{2\omega_{RF,1}}\simeq -110\,dB_m$  and $HD_{3\omega_{RF,1}}\simeq -120\,dB_m$, leading to $OIP_2\simeq 44\,dB_m$ and $OIP_3\simeq 25\,dB_m$. These new values allow to ease the design of the rest of the receiver chain, while satisfying the overall specifications.} \\
\indent {As reported in Section \ref{sec:dithering} despite the remarkable reduction of the impact of the nonlinearities due to RB, the use of the dithering tone must be controlled properly in order to avoid further distortion introduced by the dithering tone itself. In particular, Figures \ref{fig:HD2_3D}, \ref{fig:HD3_3D} and \ref{fig:IMD_dith_3D} represent experimentally the concept exposed in Section \ref{sec:dithering} with the simulations, confirming the optimum region of $10\log_{10}(M_{dith})\in[40-50]$ when $M(\omega_{RF,1})\leq 2$.}

\begin{figure}[hbpt]
\centering
\includegraphics[scale=0.4]{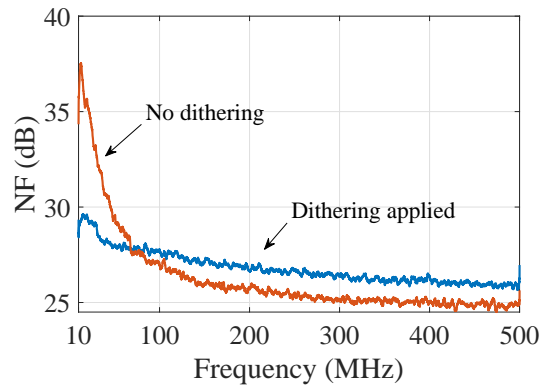}
\caption{{Comparison of the measured Noise Figure (NF) for the RoF link analyzed with and without the use of dithering tone.}}
\label{fig:Noise_Figure}
\end{figure}

{In that region, the value of $IMD_{\omega_{RF,1}+\omega_{dith}}$ falls from about $-70\,dB_m$ for the most critical point (i.e. $10\log_{10}(M_{dith})\simeq 30$) to values always lower than $-80/-90\,dB_m$, which for the system under study is well below the thermal noise at the output of the receiver considering the finest bandwidth employed, which, in case of SKA-LOW, is about $-70\,dB_m$.\\
\indent As a final consideration on the use of this technique to mitigate the effect of RB, an evaluation of the Noise Figure (NF) of the RoF link has been performed, by switching off the generators SG1 and SG2  and switching on and off the generator SGD. Figure \ref{fig:Noise_Figure} shows the experimental results obtained, comparing the case where no dithering is applied (i.e. SGD off) with the one where a dithering tone of 10 KHz with $I_{dith,0}=1.25$ $mA$ is inserted (i.e. SGD on). Based on the considerations of the previous paragraphs, this value of current leads to $10\log_{10}M_{\omega_{dith}}\simeq 47$, which falls in the optimum region of choice of $M_{dith}$ according to Figs. \ref{fig:HD2_3D_simulations}, \ref{fig:HD3_3D_simulations}, \ref{fig:IMD_dith_3D_simulations}, and \ref{fig:HD2_3D}, \ref{fig:HD3_3D}, \ref{fig:IMD_dith_3D}.}

\indent {From Figure \ref{fig:Noise_Figure} is possible to see that for very low frequency, i.e. below 60MHz, the insertion of dithering reduces the noise figure up to 5dB around 10MHz. This is due to the fact that, besides the spurious nonlinearities investigated in this work, RB generates also low frequency noise \cite{Feng,Wu} which can be mitigated using the dithering technique \cite{Pepeljugoski,Lazaro} as well as the spurious terms.\\
\indent For frequencies higher than 60MHz, a decrease in the order of 1-2 dB is observed, which typically, and especially for the application under study, can be regarded as acceptable.}

\section{Conclusion}
The possible creation of undesired nonlinear distortion terms induced by Rayleigh Backscattering in directly modulated Radio over Fiber links has been put into evidence for the first time.  A simulation program based on a rigorous mathematical model has been developed to characterize  both theoretically and experimentally the phenomenon, whose impact can be of particular importance within contexts typical of Radioastronomic Applications. A possible solution has been proposed which showed  to counteract the nonlinear behavior described, and that is at the same time of straightforward realizability.  
\bibliographystyle{IEEEtran}
\bibliography{Biblio_Rayleigh}
\end{document}